\documentclass[10pt,journal,twoside]{IEEEtran}
\usepackage{multirow}
\usepackage{amssymb}
\usepackage[dvips]{graphicx}
\usepackage{amsmath}
\usepackage{amsthm}
\usepackage{latexsym,bm}
\usepackage{color}
\usepackage{subfigure}
\usepackage{longtable}
\usepackage{accents}
\usepackage{cite}
\usepackage{enumerate}
\usepackage{arydshln}
\usepackage{slashbox}
\usepackage{algorithmic}
\usepackage{float}

\def\bB{{\mathbf{B}}}

\makeatletter
\def\widebar{\accentset{{\cc@style\underline{\mskip10mu}}}}
\def\Widebar{\accentset{{\cc@style\underline{\mskip8mu}}}}
\makeatother

\theoremstyle{plain}

\theoremstyle{definition}
\theoremstyle{definition} 
\renewcommand{\IEEEQED}{\IEEEQEDopen}
\begin{document}

\title{\huge Analysis and Optimization of Tail-Biting Spatially Coupled Protograph LDPC Codes for BICM-ID Systems
\thanks{Z.~Yang and Y.~Fang are with the School of Information Engineering, Guangdong
University of Technology, China~(email: fangyi@gdut.edu.cn).}
\thanks{G.~Zhang is with China Academy of Space Technology (Xi'an), China~(email: zhangghcast@163.com).}
\thanks{F.~C.~M.~Lau is with the Department of Electronic and Information Engineering, Hong Kong Polytechnic University, Hong Kong~(email: francis-cm.lau@polyu.edu.hk).}
\thanks{S. Mumtaz is with the Instituto de Telecomunica\c{c}\~{o}es, Portugal~(email: smumtaz@av.it.pt).}
\thanks{D. B. da Costa is with the Department of Computer Engineering, Federal University of Cear\'{a}, Sobral, CE, Brazil (email: danielbcosta@ieee.org).}}
\author{\fontsize{11pt}{\baselineskip}\selectfont {Zhaojie Yang, Yi Fang, Guohua Zhang, Francis C. M. Lau, Shahid Mumtaz, and Daniel B. da Costa}}

\maketitle

\begin{abstract}
As a typical example of bandwidth-efficient techniques, bit-interleaved coded modulation with iterative decoding (BICM-ID) provides desirable spectral efficiencies in various wireless communication scenarios.
In this paper, we carry out a comprehensive investigation on tail-biting (TB) spatially coupled protograph (SC-P) low-density parity-check (LDPC) codes in BICM-ID systems. Specifically, we first develop a two-step design method to formulate a novel type of constellation mappers, referred to as {\em labeling-bit-partial-match (LBPM) constellation mappers}, for SC-P-based BICM-ID systems. The LBPM constellation mappers can be seamlessly combined with high-order modulations, such as $M$-ary phase-shift keying (PSK) and $M$-ary quadrature amplitude modulation (QAM).
Furthermore, we conceive a new bit-level interleaving scheme, referred to as {\em variable node matched mapping (VNMM) scheme}, which can substantially exploit the structure feature of SC-P codes and the unequal protection-degree property of labeling bits to trigger the wave-like convergence for TB-SC-P codes.
In addition, we propose a hierarchical extrinsic information transfer (EXIT) algorithm to predict the convergence performance (i.e., decoding thresholds) of the proposed SC-P-based BICM-ID systems.
Theoretical analyses and simulation results illustrate that the LBPM-mapped SC-P-based BICM-ID systems are remarkably superior to the state-of-the-art mapped counterparts. Moreover, the proposed SC-P-based BICM-ID systems can achieve even better error performance with the aid of the VNMM scheme. As a consequence, the proposed LBPM constellation mappers and VNMM scheme make the SC-P-based BICM-ID systems
a favorable choice for the future-generation wireless communication systems.
\end{abstract}


\section{Introduction}\label{sect:Introduction}

\begin{figure*}[!htpb]
\centering\vspace{-2mm}
\includegraphics[width=4.6in,height=1.7in]{{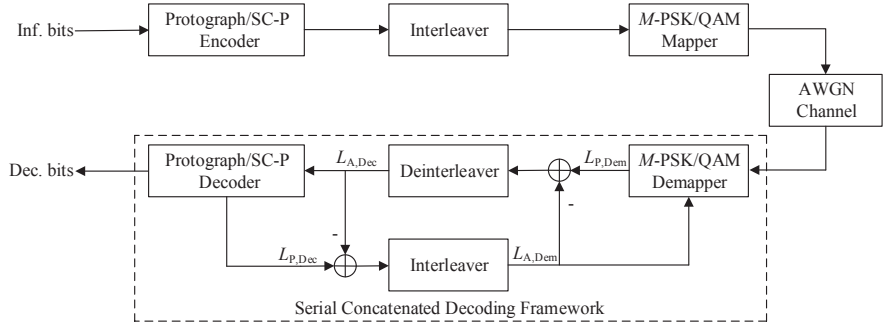}}
\caption{Block diagram of a protograph/SC-P-based BICM-ID system over an AWGN channel.}
\label{fig:BICM-ID}  
\end{figure*}

As a type of bandwidth-efficient coded modulation techniques, bit-interleaved coded modulation (BICM) has attracted a significant amount of attention in the past two decades. BICM is constructed by a serial concatenated framework, which is composed of a channel encoder, a bit interleaver, and a constellation mapper \cite{669123}. This serial concatenated framework not only can boost the flexibility of system design, but also can improve the system performance and throughput in various channel environments.
However, the BICM is regarded as a sub-optimal transmission scheme because of the independent demapping procedure.
Recently, some signal shaping techniques have been widely investigated to improve the BICM capacity \cite{7419222}. As one of the most popular shaping techniques, the probabilistic shaping technique allows the constellation labels to have non-equiprobable probabilities and thus realizes desirable shaping improvement \cite{7307154}.
Furthermore, an iterative processing for soft information between the demapper and the channel decoder has been employed at the receiver to improve the system performance. This operation results in an enhanced version of BICM systems, referred to as {\em BICM with iterative decoding (BICM-ID) systems} \cite{1025496}. Such systems not only can inherit the high spectral-efficiency advantage, but also can exhibit better error performance.

Thanks to the aforementioned advantages, a large set of research activities have been sparked off to analyze and optimize the BICM-ID systems in order to enhance the reliability and rate of data transmission. For instance, many researches have focused on the design of constellation mapper, which is one critical parameter that determines the system performance.
It is noteworthy that a constellation mapper that exhibits excellent performance in BICM systems is not always suitable for BICM-ID systems due to the incorporation of the iterative decoding technique.
As a typical example, Gray constellation mapper is considered as an optimal constellation mapper in BICM systems, but this mapper can only achieve trivial performance gain when ID is introduced \cite{6612622,6570510,6809890}. For this reason, some constellation mappers have been designed particularly for BICM-ID systems. In \cite{6823707}, an innovative method, which can be used to generate good phase-shift keying (PSK) and quadrature amplitude modulation (QAM) constellation mappers for BICM-ID systems, has been introduced. Moreover, a constellation-design criterion has been proposed to enhance the performance of BICM-ID systems over two-way relay fading channels \cite{8052124,7447812}. In \cite{8053803}, the authors have proposed a systematic method to construct multi-dimensional (MD) mappings based on the $16$- and $64$-QAM constellations.

Besides depending on the constellation mappers, the performance of BICM-ID systems depends on the error-correction codes (ECCs). Thus, much research effort has been endeavored to promote the analysis and optimization of ECCs for BICM-ID systems. Among all ECCs, protograph low-density parity-check (LDPC) codes are the most popular due to their excellent error performance and relatively low complexity \cite{7112076}. During the past decade, much attention has been paid to developing theoretical-analysis tools for protograph codes so as to determine the best candidates for a specific channel environment. As an example, the extrinsic information transfer (EXIT) algorithm has been conceived to estimate the asymptotic-performance metrics of protograph codes (i.e., the decoding thresholds) \cite{4411526}. In recent years, the combination of protograph codes and BICM technique has been carefully investigated in bandwidth-limited communication environments.
To be specific, an interleaving scheme, called {\em variable degree matched mapping (VDMM) scheme}, has been proposed and optimized in order to improve the performance of protograph-based BICM systems \cite{1605713,5613828,6133952,6777400}. In parallel with the above advancement, the optimization of protograph-based BICM-ID systems has also been studied over Poisson pulse-position modulation (PPM) channels \cite{6663748,6955112}.


In addition to the protograph codes, a type of LDPC convolutional-like codes, referred to as {\em spatially coupled (SC) LDPC codes}, has gained increasing interest in the coding communit \cite{0011223,1362891}. Compared with block codes, the SC codes are capable of achieving better performance because of the convolutional gain \cite{5571910,5695133,6589171,5695130}.
Thus, there is a surge of research interest in developing some practical design methods for SC codes to facilitate their application in wireless communication scenarios.
For instance, the time-invariant SC codes with small constraint length \cite{8114265,8580758,8339512} and the finite-length SC codes \cite{8272426,8449217}, which benefit from desirable error performance and decoding complexity, have been deeply investigated in recent years.
In parallel with the above advancements, some energy allocation and high-order modulation techniques have been considered for SC codes.
In particular, an energy shaping scheme, which allows the system to allocate different transmission energies for different coded bits, has been proposed to improve the decoding thresholds of tail-biting SC (TB-SC) codes over additive white Gaussian noise (AWGN) channels \cite{8281448}.
Moreover, some improved bit-mapping schemes have been proposed for TB-SC codes in BICM frameworks so as to accelerate the convergence of the decoding process \cite{7005396,7460483}.
Besides, the idea of EXIT has been exploited to optimize the asymptotic performance of SC codes in different transmission scenarios \cite{8108160,8377289,7593089}.

Inspired by the appealing advantages of protograph codes and SC codes, spatially coupled protograph (SC-P) codes have also been formulated. The SC-P codes can not only inherit all superiorities of protograph codes, but also possess the convolutional property to achieve additional performance gain \cite{6262475,7152893}. Accordingly, such codes have become a more preferable choice in practical applications. During the past five years, abundant works have been reported to promote the theoretical and practical advancement of SC-P codes in different transmission environments, such as binary erasure channels (BECs) and AWGN channels \cite{7339427,7353121}. Nevertheless, to the best of our knowledge, the performance analysis and optimization of SC-P codes for BICM systems, especially BICM-ID systems, are relatively unexplored.

With the above motivation, this paper conducts an insightful and comprehensive study on tail-biting SC-P (TB-SC-P) codes for BICM-ID systems.
The main contributions of this paper can be summarized as follows. First, according to the average mutual information (AMI) analysis, a two-step design method is proposed to construct a novel type of constellation mappers, referred to as {\em labeling-bit-partial-match (LBPM) constellation mappers}, for the SC-P-based BICM-ID systems. An appealing benefit of the proposed constellation mappers is that they can not only achieve higher capacity than other existing constellation mappers in BICM scenarios, but also accomplish larger iterative improvement in BICM-ID scenarios.
Moreover, we conceive a novel interleaving scheme, i.e., {\em variable node matched mapping (VNMM) scheme}, tailored for the TB-SC-P codes so as to offer more performance gain in the BICM-ID scenarios. Especially, the proposed VNMM scheme can give rise to the decoding-wave phenomenon in the joint demapping-and-decoding process and thus significantly accelerate the convergence of the a posteriori MI of TB-SC-P codes.
Additionally, a hierarchical EXIT algorithm, which can be used to evaluate the asymptotic convergence performance of TB-SC-P codes with different constellation mappers, is proposed for BICM-ID systems.
To summarize, the proposed LBPM-VNMM-aided SC-P-based BICM-ID system is superior to the state-of-the-art BICM-ID systems and stands out as a promising transmission technique for future wireless communication systems.

The remainder of this paper is organized as follows. Sect.~II presents the SC-P-based BICM-ID system model and its AMI analysis. In Sect.~III, we put forward a two-step design method to construct the LBPM constellation mappers and we illustrate its merit from the information-theoretical perspective. In Sect.~IV, we further conceive a VNMM scheme for the SC-P-based BICM-ID systems. In Sect.~V, we propose a hierarchical EXIT algorithm to derive the decoding thresholds of TB-SC-P codes in the BICM-ID systems. Sect.~VI gives various simulation results along with several discussions and insights, and Sect.~VII concludes the overall work.

\section{System Model and AMI Calculation}\label{sect:section-2}

In this section, we begin with a brief introduction of the SC-P-based BICM-ID systems over an AWGN channel.
Then, we review the information-theoretical analysis of constellation-constrained AMI for such systems.

\subsection{BICM-ID Systems}\label{sect:section-2-A}
The protograph/SC-P-based BICM-ID system considered in this paper is illustrated in Fig.~\ref{fig:BICM-ID}. As shown, the information bits (Inf. bits) are first encoded by a protograph/SC-P encoder to yield a coded bit sequence. Such a sequence is permuted by a bit-level interleaver. Then, every $m$ bits output from the interleaver are mapped to a modulated symbol based on a specific $M$-ary constellation mapper, such as $M$-PSK and $M$-QAM constellation mappers. After the above process, a coded bit sequence of length $n$ is converted to an $M$-ary modulated symbol sequence of length $n/m$, where $m=\log_2M~(m=3,4,\ldots)$.\footnote{In this paper, we focus on the modulations with $m>2$ because the proposed LBPM constellation mapper reduces to an anti-Gray constellation mapper when $m=2$. Interested readers are referred to Sect.~III for detailed principles of the proposed LBPM constellation mappers.} The $j$-th modulated signal $x_j$ will be transmitted through a complex AWGN channel and its corresponding received signal $y_j$ is given by
\begin{equation}
y_j= x_j + n_j,
\label{eq:eq1}
\end{equation}
where $j = 1,2,\ldots,n/m$; $x_j$ belongs to a constellation set $\chi$; $|\chi| = M = 2^m$ denotes the size of constellation set; $n_j$ represents the complex Gaussian noise with zero mean and variance $\sigma^2 = N_0/2$ in each dimension. In this paper, it is assumed that the transmitted energy per symbol is normalized to one, i.e., $E_s= {\mathbb E} [|x_j|^2] = 1$, where $\mathbb{E}[\cdot]$ denotes the expectation function.

At the receiver, the received signal will be processed by a serial concatenated decoding framework, which consists of a single-input single-output (SISO) demapper and a SISO decoder.
Specifically, based on the received signal $y_j$ and the a-priori log-likelihood ratios (LLRs) of the demapper, the extrinsic LLRs output from the demapper can be computed and sent to the deinterleaver.
After a specific deinterleaving operation, these extrinsic LLRs will be used as a-priori LLRs of the decoder.
Subsequently, the extrinsic LLRs output from the decoder can be calculated and sent to the interleaver.
After being permuted by the interleaver, these extrinsic LLRs will be fed back to the demapper and used as the updated a-priori LLRs of the demapper for the next iteration.
Based on such a framework, the extrinsic LLRs can be updated iteratively to accelerate the convergence of a-posteriori LLRs and to enhance the performance of BICM-ID systems.
In this work, the max-sum approximation of the log-domain maximum a-posteriori probability (Max-Log-Map) algorithm \cite{6942236} and belief-propagation (BP) algorithm \cite{7152893} are utilized to implement the demapper and decoder, respectively.\footnote{Although the standard BP algorithm is utilized to decode the SC-P codes in this paper, the windowed BP algorithm \cite{8281448} is also applicable to decoding the proposed BICM-ID systems.}


\subsection{Constellation-Constrained AMI Calculation}\label{sect:section-2-B}
It is well known that the AMI between the channel input and the channel output determines the maximum information rate for error-free transmission \cite{8237200,8186234}.
In this sense, the AMI can be treated as a criterion for estimating the channel transmission performance.
Suppose that the transmitted signal is taken from a given constellation set with equal probability,
the constellation-constrained AMI of CM over an AWGN channel can be evaluated as
\begin{equation}
C_{\rm CM} = m - \mathbb{E}_{x,y} \left[ \log_{2} \frac{\sum\nolimits_{z\in\chi}p(y|z)}{p(y|x)} \right],
\label{eq:eq2}
\end{equation}
where $p(y|z)$ denotes the probability density function (PDF) of the received signal $y$ conditioned on the complex symbol $z$ chosen from a constellation set $\chi$, and $p(y|x)$ denotes the PDF of the received signal $y$ conditioned on the modulated signal $x$. The $C_{\rm CM}$ defined in (\ref{eq:eq2}) is called the {\em coded modulation (CM) capacity}. According to \cite{5073425,3698521}, CM can maximize the achievable rate and thus is regarded as an optimal transmission scheme.

On the other hand, for a BICM system, the constellation-constrained AMI can be measured by
\begin{equation}
C_{\rm BICM} = m - \sum\limits_{i=1}^{m}\mathbb{E}_{b,y} \left[ \log_{2} \frac{\sum\nolimits_{z\in\chi}p(y|z)}{{\sum\nolimits_{z\in\chi_{i}^{(b)}}}{p(y|z)}} \right],
\label{eq:eq3}
\end{equation}
where $\chi_{i}^{(b)}$ denotes the subset of constellation set $\chi$ with the $i$-th bit being $b$ $(b\in\{0,1\})$.
The $C_{\rm BICM}$ defined in (\ref{eq:eq3}) is called the {\em BICM capacity}.

As is well known, BICM is regarded as a sub-optimal transmission scheme and is not able to achieve the maximum achievable rate due to the independent demapping
procedure\cite{7095513}. To address this issue, an iterative processing for soft information exchange between the demapper
and decoder can be employed at the receiver of a BICM to formulate a BICM-ID. In a BICM-ID system, the a-priori information of $m$ labeling bits within a modulated symbol is fed back from the decoder to the demapper. This operation can be exploited substantially to improve the overall performance (i.e., boost the achievable rate) of the BICM system. After a sufficient number of iterations between the demapper and decoder, BICM-ID can perfectly compensate the rate loss of BICM and hence maximize the achievable rate \cite{1025496,7095513,1478523}. For the above reason, BICM-ID can be considered as an optimal transmission scheme whose constellation-constrained AMI is equal to the CM capacity \cite{5073425,8527419}.
In this sense, $C_{\rm CM}$ can be treated as the {\em BICM-ID capacity}, which is denoted as $C_{\rm BICM-ID}$.

\section{Design and Analysis of LBPM Constellation Mappers for SC-P-based BICM-ID Systems}\label{sect:section-3}
\subsection{Proposed Design Method}
It is well known that the constellation mapper is one of the most critical parameters for bandwidth-efficient transmission techniques \cite{6570510}.
In different communication environments, the constellation mapper should be carefully designed and optimized so as to improve the system performance.
In fact, any constellation mapper has its specific application scenarios.
For instance, the Gray constellation mapper can achieve optimal performance in the BICM systems \cite{6570510,6809890}.
But it is not suitable for BICM-ID systems owing to the trivial iterative gain \cite{6612622,6570510,6809890}.
To address this problem, a variety of constellation mappers, which can obtain larger iterative gain than the Gray constellation mapper in the BICM-ID systems, have been proposed \cite{1413232,6570510,8052124,7935441}.
Fig.~\ref{fig:side:2} shows three classical constellation mappers, i.e., set-partition (SP), maximum squared Euclidean weight (MSEW), and anti-Gray mappers for the $8$-PSK and $16$-QAM modulations, which are suitable for BICM-ID scenarios.
In this figure, $x_l$ is the $l$-th label in the constellation (i.e., $l=1,2,\ldots,2^m$) and $x^{b_i}_l$ denotes the $i$-th labeling bit within $x_l$ (i.e., $i=1,2,\ldots,m$).

From the perspective of equivalent parallel channels, an $M$-ary modulation scheme can be viewed as a set of $m$ parallel independent and memoryless binary input sub-channels (i.e., $m=\log_2M$) \cite{669123}.
Such $m$ sub-channels have different reliabilities (i.e., different AMIs corresponding to the $m$ sub-channels) and each sub-channel corresponds to a labeling bit. In this paper, we define the protection degree of a labeling bit as the reliability of its corresponding sub-channel.
Considering a $2^m$-ary modulation scheme, there are $2^m$ labels in the corresponding constellation mapper and each label is composed of $m$ labeling bits.
From the distribution of labels in the constellation, it is apparent that the protection degrees for the $m$ labeling bits are not exactly identical.
Besides, it has been demonstrated in \cite{8331859} that the Hamming distance of adjacent labels has a significant influence on the system performance.
Based on the above discussion,
a two-step design method, which can be used to construct the LBPM constellation mappers for high-order modulations, is proposed as follows.

\begin{enumerate}
\item
{\bf Maximizing the number of highly protected labeling bits:}
For a $2^m$-PSK or QAM constellation mapper, there exist $2^m$ labels and each label contains $m$ labeling bits.
The protection degrees corresponding to the $m$ labeling bits can be evaluated by analyzing all the labeling-bit distribution in the constellation mapper.
The higher the protection degree of a labeling bit is, the greater the AMI between the labeling bit and the received signal can be acquired.
In this work, we refer to these labeling bits with maximum protection degree as \textit{highly protected labeling bits} and the rest labeling bits as \textit{generally protected labeling bits}.
With an aim to maximize the AMI, the number of highly protected labeling bits should be maximized.
Specifically, suppose that the labeling bit sequence corresponding to the label $x_l$ is represented by $\{x_l^{b_1},x_l^{b_2},\ldots,x_l^{b_m}\}$ and the number of highly protected labeling bits is assumed to be $m'$. The parameter $m'$, which satisfies $0<m'<m$, can be calculated based on the labeling bit distribution in a constellation mapper.
Afterwards, we can place the highly protected labeling bits of the label $x_l$ at the first $m'$ positions (i.e., $\{x_l^{b_1},x_l^{b_2},\ldots,x_l^{b_{m'}}\}$) within the labeling bit sequence.
The above {\em labeling-bit partial matched} operation can ensure that the number of highly protected labeling bits can be maximized and thus can make the BICM-ID systems having excellent performance even without ID.

\item
{\bf Maximizing the Hamming distance between the adjacent labels:}
To boost the iterative gain in the context of ID,
we propose a {\em maximizing-Hamming-distance} criterion to deal with the remaining $m-m'$ generally protected labeling bits in the label $x_l$.
To begin with,
the first generally protected labeling bit is placed at the $m'+1$ position in $x_l$.
This bit should be reasonably set to maximize the Hamming distance between the adjacent labels in the current constellation.
Likewise,
the remaining generally protected labeling bits, which are respectively placed at the $m'+2,m'+3,\ldots,m$ positions in the labeling bit sequence, will be processed individually according to the proposed maximizing-Hamming-distance criterion.
As a consequence, an LBPM constellation mapper can be successfully constructed.
\end{enumerate}

\begin{figure}[tbp]
\centering
\subfigure[\hspace{0.07cm}]{ 
\includegraphics[width=1.6in,height=1.53in]{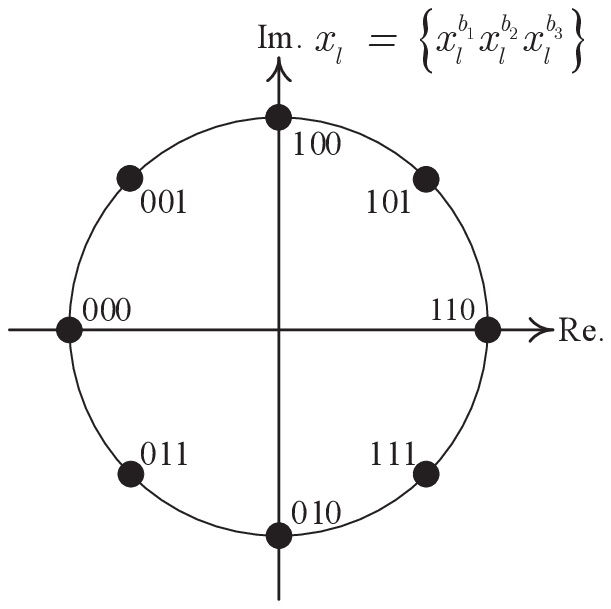}}
\subfigure[\hspace{0.22cm}]{ 
\includegraphics[width=1.6in,height=1.53in]{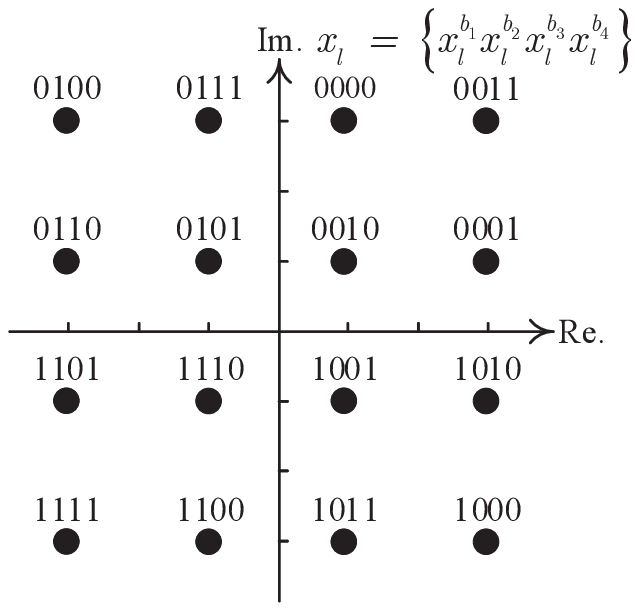}}
\caption{Constellation mappers of the proposed LBPM: (a) $8$-PSK and (b) $16$-QAM modulations.}
\label{fig:side:3}  
\end{figure}

Following the proposed design method, one can construct the corresponding LBPM constellation mappers for different modulations.
As an example, we show the LBPM constellation mappers for $8$-PSK and $16$-QAM modulations in Fig.~\ref{fig:side:3}.
In this figure, we can see that the labeling bits $x_l^{b_1}$ and $x_l^{b_2}$ belong to the highly protected labeling bits for both the $8$-PSK and $16$-QAM constellation mappers.
However, the remaining labeling bits (i.e., $x_l^{b_3}$ for $8$-PSK while $x_l^{b_3}$ and $x_l^{b_4}$ for $16$-QAM), which belong to the generally protected labeling bits, are carefully set to maximize the Hamming distance between adjacent labels.
In order to verify the superiority of our proposed LBPM constellation mappers in the BICM framework, we will analyze the BICM capacities by exploiting such mappers for $8$-PSK and $16$-QAM in the next subsection. As benchmarks,  three state-of-the-art constellation mappers, i.e., SP, MSEW, and anti-Gray, are also considered. The benefit of our proposed LBPM constellation mappers in the BICM-ID framework will be further illustrated in the forthcoming sections. In particular, Sect.~III-B (i.e., Fig.~\ref{fig:side:a}) and Sect.~V-B (i.e., Table~\ref{tab:tab1}) verify the effectiveness of Step 1) and Step 2) of the proposed LBPM constellation mappers, respectively.

\subsection{Capacity Analysis}
\begin{figure}[tbp]
\centering\vspace{-3mm}
\subfigure[\hspace{-0.5cm}]{ 
\includegraphics[width=3.0in,height=2.3in]{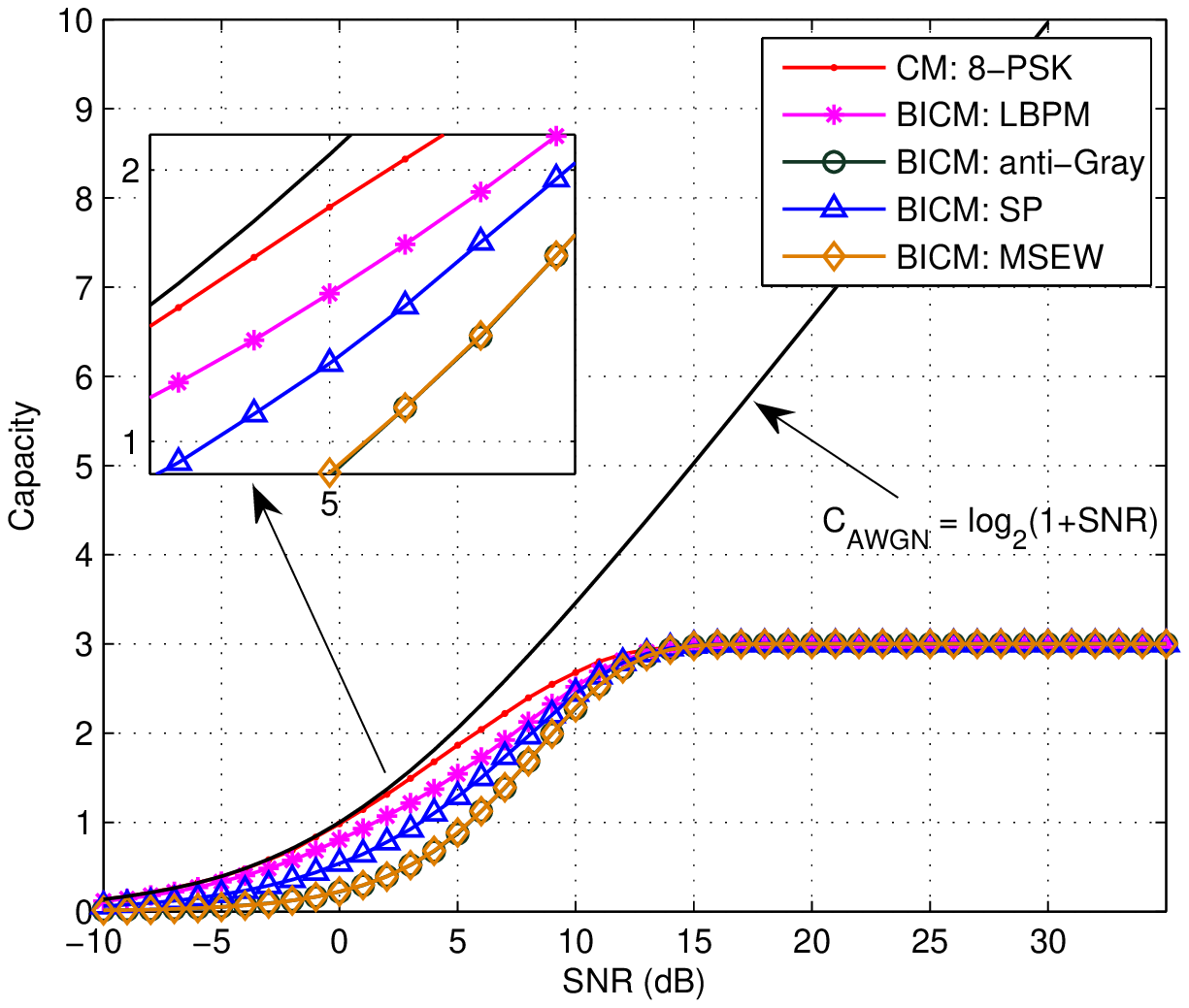}}
\subfigure[\hspace{-0.5cm}]{ 
\includegraphics[width=3.0in,height=2.3in]{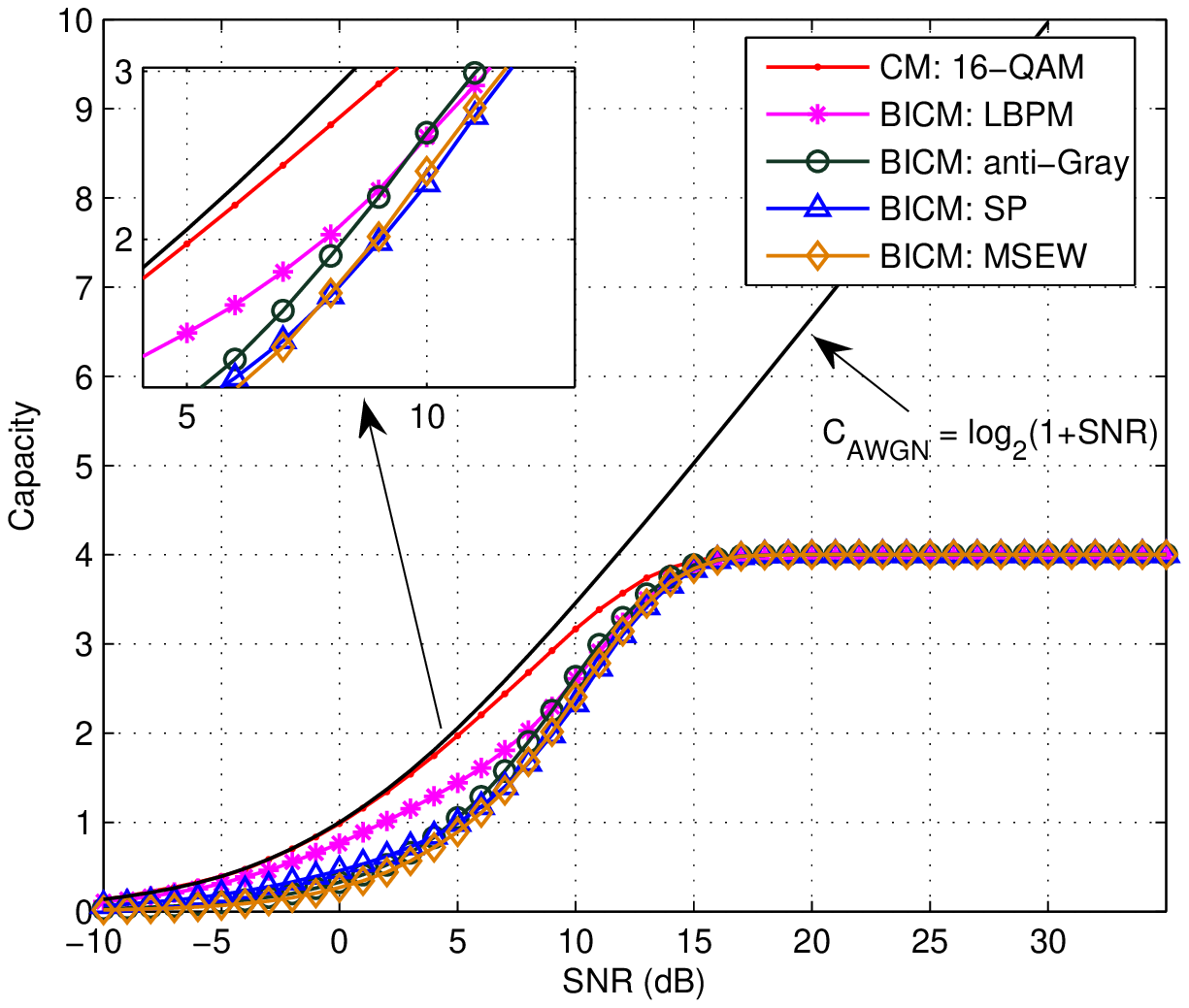}}
\caption{The CM and BICM capacities for the proposed LBPM and three existing constellation mappers (i.e., SP, MSEW, and anti-Gray) over an AWGN channel: (a) $8$-PSK and (b) $16$-QAM modulations.}
\label{fig:side:a}  
\end{figure}

The BICM capacities of different constellation mappers can be calculated by exploiting the constellation-constrained AMI analytical method in Sect.~\ref{sect:section-2}. In the case of an AWGN channel, the CM and BICM capacities for $8$-PSK and $16$-QAM modulations with the proposed LBPM and three existing constellation mappers are depicted in Fig.~\ref{fig:side:a}.
As observed from Fig.~\ref{fig:side:a}(a),
the CM capacity is better than the BICM capacity ($C_{\rm CM} \ge C_{\rm BICM}$), indicating that the BICM is a sub-optimal transmission scheme.
More importantly, the proposed LBPM constellation mapper leads to a larger capacity with respect to the other three constellation mappers when the code rate $R \le 0.5$ (i.e., $R = C_{\rm BICM}/m$). One can also observe that the gap between the CM capacity and BICM capacity becomes the smallest when using the LBPM constellation mapper.
Therefore, it can be concluded that the BICM system with the proposed constellation mapper can achieve better performance compared with counterparts using SP, MSEW, and anti-Gray constellation mappers.
Similar conclusions can also be drawn from the case of $16$-QAM modulation (see Fig.~\ref{fig:side:a}(b)).

\section{Interleaver Design for SC-P-based BICM-ID Systems}\label{sect:section-4}

\subsection{Terminated SC-P Codes}\label{sect:section-4-A}

A protograph, which has been proposed in \cite{7112076,Thorpe2003Low}, can be represented by a small Tanner graph \cite{Thorpe2003Low} ${\cal G = (V, C, E)}$ consisting of three different sets, i.e., a variable-node (VN) set ${\cal V}$, a check-node (CN) set ${\cal C}$, and an edge set ${\cal E}$.
The cardinalities of ${\cal V}$, ${\cal C}$, and ${\cal E}$ equal $n_p$, $m_p$, and $E$, respectively.
For a protograph, each edge $e_{i,j} \in {\cal E}$ connects a VN $v_j \in {\cal V}$ with a CN $c_i \in {\cal C}$, and  parallel edges are allowed since they can be thoroughly eliminated during an expansion procedure.
Furthermore, a protograph with a code rate $R = (n_p - m_p)/n_p$ can be defined by a base matrix $\bB = (b_{i,j})$ of size $m_p \times n_p$, where $b_{i,j}$ denotes the number of edges connecting $c_i$ with $v_j$. 
An expanded protograph (resp. parity-check matrix) corresponding to a protograph code can be generated by performing the ``copy-and-permute" operation on a given protograph (resp. base matrix) \cite{Thorpe2003Low}.
In actual implementation, the ``copy-and-permute'' operation (i.e., also known as ``lifting'' operation) can be realized by exploiting a modified progressive-edge-growth (PEG) algorithm \cite{van2012design}.
After that, a protograph code of a given length can be easily constructed based on its parity-check matrix.

{\em Remark:} In this paper, both the protograph codes and TB-SC-P codes are generated by performing the modified PEG algorithm on their corresponding protographs.

By replicating a protograph $L$ times and associating each protograph with a time index $t$ $(t=0, 1, \ldots, L-1)$,
one can construct an SC-P code by spatially coupling the sequence of $L$ disjoint protograph codes into a single coupled chain, where $L$ is the coupling length.
More specifically, the edges emanating from the VNs of protograph at times $t$ $(t=0, 1, \ldots, L-1)$ are connected to the CNs of protograph at times $t, t+1, \ldots, t+w$ based on a given edge-spreading rule, where $w$ denotes the coupling width ($0< w < L$).
From the perspective of matrix structure, given a specific edge-spreading rule and a coupling width $w$,
the base matrix $\bB$ corresponding to a protograph of size $m_p \times n_p$ can be decomposed into $w+1$ submatrices (i.e., $\bB_0, \bB_1, \cdots, \bB_w$).
Moreover, the relationship between the base matrix and submatrices is governed by $\sum_{k=0}^w\bB_k=\bB$ and all matrices must be of the same size.
Based on the above discussion, the base matrix corresponding to a terminated SC-P (TE-SC-P) code can be expressed as
\vspace{-1.5mm}
\begin{equation}
\bB_{L}^{\rm TE}=
  \begin{array}{c}
                      L     \\
\begin{bmatrix}
 \overbrace{
  \begin{array}{ccccccccccccc}
    \bB_0 &        &       &         &         &       &         &\\
    \bB_1 & \bB_0  &       &         &         &       &         &\\
  \vdots  & \bB_1  & \ddots&         &         &       &         &\\
    \bB_w & \vdots & \ddots&  \bB_0  &         &       &         &\\
          & \bB_w  &       &  \bB_1  &  \bB_0  &       &         &\\
          &        & \ddots&  \vdots &  \bB_1  & \ddots&         &\\
          &        &       &  \bB_w  &  \vdots & \ddots&  \bB_0  &\\
          \hdashline
          &        &       &         &  \bB_w  &       &  \bB_1  &\\
          &        &       &         &         & \ddots&  \vdots &\\
          &        &       &         &         &       &  \bB_w  &\\
 \end{array}}
\end{bmatrix}\label{eq:TE}
  \end{array},
\end{equation}
where the size of $\bB_L^{\rm TE}$ is equal to ${m_p (L+w) \times n_p L}$.

For a finite coupling length $L$, the code rate $R_L^{\rm TE}$ corresponding to a TE-SC-P code can be measured by
\begin{equation}
\begin{split}
R_L^{\rm TE}=1-\frac{(L+w)m_p}{Ln_p}=1-\left(\frac{L+w}{L}\right)(1-R)
\end{split}.\label{eq:R-TE}
\end{equation}

It is apparent that the code rate $R_L^{\rm TE}$ is less than $R$, i.e., $R_L^{\rm TE} < R$.
To overcome the rate-loss weakness, another termination method has been developed to formulate a new type of SC-P codes, called {\em TB-SC-P codes}.

\subsection{Tail-Biting SC-P Codes}\label{sect:section-4-B}
A TB-SC-P code can be generated from a TE-SC-P code by combining the CNs at times $t=L,L+1,\dots,L+w-1$ with the corresponding CNs of the same type at times $t = 0,1,\dots,w-1$, respectively. Consequently, the base matrix corresponding to a TB-SC-P code can be given by
\begin{equation}
\bB_L^{\rm TB}=
\begin{bmatrix}
  \begin{array}{llllll}
    \bB_0 &          &           &  \bB_w     &  \cdots & \bB_1    \\
 \vdots   &   \ddots &           &            &  \ddots & \vdots   \\
 \bB_{w-1}&          &           &            &         & \bB_w    \\
    \bB_w &          &  \bB_0    &            &         &          \\
          &   \ddots &  \vdots   &  \bB_0     &         &          \\
          &   \ddots &  \bB_{w-1}&  \vdots    &  \ddots &          \\
          &          &  \bB_w    &  \bB_{w-1} &  \cdots & \bB_0    \\
  \end{array}
\end{bmatrix},\label{eq:TB}
\end{equation}
where the size of $\bB_L^{\rm TB}$ equals ${m_p L \times n_p L }$.
By comparing \eqref{eq:TB} and \eqref{eq:TE},
one can easily find that the base matrix $\bB_L^{\rm TB}$ can be obtained from the base matrix $\bB_L^{\rm TE}$ via adding its last $m_p w$ rows to the first $m_p w$ rows.
As a result, the TB-SC-P codes have the same code rate and degree distribution as their corresponding protograph codes.

\subsection{Interleaver Design}\label{sect:section-C}
\subsubsection{Related Work}
It is well known that the TE-SC-P codes can attain a larger threshold improvement as the coupling length $L$ tends to infinity.
Based on the information-theoretical analysis, this threshold improvement is caused by the decoding-wave phenomenon in the decoding process, which can be explained as follows \cite{5695130,7152893,7339427}.
Considering a TE-SC-P code, the MIs corresponding to the VNs at both ends of the protograph can converge with relatively higher priorities compared with the remaining VNs.
This will further help to accelerate the convergence speed of MIs corresponding to the remaining VNs during the iteration procedure.
The decoding-wave phenomenon greatly reduces the SNR required for the convergence of MIs of all VNs and thus improves the decoding threshold of the TE-SC-P code.
Nonetheless, a large coupling length will greatly increase the complexity of TE-SC-P codes, and hence makes it rather difficult to meet the low-latency requirement for practical communication systems.
When the coupling length $L$ is small or moderate, there exists a rate-loss problem in such codes and the transmission-rate requirement can not be met.

Compared with the TE-SC-P codes, the TB-SC-P codes do not have the threshold improvement, but they have the same code rates as the original protograph codes.
In order to improve the decoding thresholds of such codes, a hybrid mapping scheme, which simultaneously adopts two constellation mappers for each codeword, has been proposed in \cite{7593089}. However, the hybrid mapping scheme may suffer from higher implementation complexity. Moreover, the work has only considered the $16$-QAM modulation in the mapping design.
In \cite{7460483}, a shortening technique has been applied for TB-SC codes.
The principle is that one can shorten the codeword by setting some coded bits to zero and thus inject some priori knowledge. 
But such a technique is at the cost of sacrificing the code rate.
Based on the unequal protection-degree property of the labeling bits in SC-P-based BICM-ID systems (see Sect.~\ref{sect:section-3}),
we propose a new interleaving scheme to protect some coded bits in the bit-to-symbol-mapping process. The new interleaving scheme not only can maintain the same implementation complexity and code rate as those of original systems, but also can trigger the decoding-wave phenomenon for TB-SC-P codes so as to improve their decoding thresholds in the BICM-ID systems. In the following, we will briefly describe the principle of the new interleaving scheme, referred to as {\em VNMM scheme}, for TB-SC-P-based BICM-ID systems.


\subsubsection{Proposed VNMM Scheme}

For an $M$-ary BICM-ID system, there exist $M$ different labels. Each label consists of $m$ labeling bits. Based on the distribution of $M$ labels in the constellation mapper, one can evaluate the protection degrees for the $m$ labeling bits (i.e., the protection degrees for different labeling bits are considered as the reliabilities for different sub-channels).
Afterwards, the two labeling bits with relatively higher protection degrees than the remaining $m-2$ labeling bits can be found (i.e., $m \geq 3$) and referred as the highly protected labeling bits. However, the remaining $m-2$ labeling bits are viewed as the generally protected labeling bits.
In the bit-to-symbol mapping process, one can extract $m$ coded bits from the $m$ different blocks separately to generate a modulated $M$-ary symbol.
Specifically, the coded bits, which are extracted from the first block (i.e., the first $n/m$ coded bits) and the last block (i.e., the last $n/m$ coded bits), must be assigned to the two highly protected labeling bit positions.
On the other hand, the remaining $m-2$ coded bits can be extracted from the remaining $m-2$ blocks and assigned to the generally protected labeling bit positions in a sequential order.

After the above interleaving operation, one can assume that the coded bits within these $m$ different blocks are assigned into $m$ parallel sub-channels with different reliabilities.
For this reason, the convergence performances of AMIs for the coded bits within different blocks are not exactly the same. In particular, the AMIs corresponding to the coded bits at both ends of the codeword converge faster than that of coded bits in the middle.
Therefore, the wave-like convergence for TB-SC-P codes can be triggered in the iterative decoding scenario. The newly proposed interleaving scheme that considers the unequal-protection-degree property is referred to as the {\em VNMM scheme}.

In general, the principle of the proposed VNMM scheme for a $2^m$-ary modulation is illustrated in Fig.~\ref{fig:Interleaver}. As can be observed, the VNMM scheme can substantially exploit the structural feature of SC-P codes and the unequal protection-degree property of labeling bits so as to achieve additional performance gain in the BICM-ID systems.

{\em Remark:} The wave-like convergence principle of our proposed VNMM scheme is quite different from that of the existing techniques \cite{8281448,7005396,7460483}. In our proposed VNMM scheme, the unequal protection-degree property of the labeling bits in a given constellation mapper can be exploited to protect some specific coded bits at both ends of the spatially coupled chain with relatively higher priorities in the bit-to-symbol mapping process.
During the iterations between the demapper and decoder (i.e., outer iterations), the MIs corresponding to coded bits at both ends of the codeword can converge with relatively higher priorities and further accelerate the convergence of MIs corresponding to the remaining coded bits, which effectively triggers the decoding-wave phenomenon.


Assuming that the modulation order adopted in the BICM-ID systems is $M$, every $m$ bits are grouped together and then mapped onto a modulated $M$-ary symbol (i.e., $m=\log_2M$).
In this sense, the entire codeword can be divided into $m$ blocks in a sequential order and each block consists of $n/m$ bits.
Based on the LBPM constellation mappers,
we now give two different examples in order to clearly explain how to implement the proposed VNMM scheme.

{\bf \emph{Example 1:}} Considering a $8$-PSK modulation and a TB-SC-P code, every $m=3$ coded bits are modulated into a symbol. Then, the entire codeword can be divided into $m=3$ blocks in a sequential order, each of which contains $n/3$ coded bits.
In the proposed $8$-PSK LBPM constellation mapper, the number of highly protected labeling bits equals $2$ (i.e., the $x_l^{b_1}$ and $x_l^{b_2}$ in Fig.~\ref{fig:side:3}(a) are the highly protected labeling bits).
Assuming that the AMI between the labeling bits $x_l^{b_i}$ and the received signal $y_j$ is denoted by $I(x^{b_i};y)$,
the three labeling bits in each label must satisfy the following property
$$I(x^{b_1};y)=I(x^{b_2};y)>I(x^{b_3};y).$$
In the process of bit-to-symbol mapping, a symbol is generated by interleaving three coded bits, which can be extracted from the predefined three different blocks separately.
Specifically, the VNs at both ends of the entire codeword should be protected with relatively higher priorities in order to accelerate their MI convergence in the decoding process.
To achieve this goal, the coded bits in the first block and the last block should be assigned to the labeling bit positions $x^{b_1}_l$ and $x^{b_2}_l$, respectively.
On the contrary, the coded bits in the second block should be assigned to the remaining labeling bit position $x^{b_3}_l$. Exploiting the above VNMM scheme, the decoding-wave phenomenon of the TB-SC-P code can be successfully triggered.

{\bf \emph{Example 2:}} Likewise, for a $16$-QAM modulation, every $m=4$ coded bits are modulated into a symbol.
In this case, one can divide the whole codeword into $m=4$ blocks, where $n/4$ coded bits are involved in each block.
Referring to the proposed $16$-QAM LBPM constellation mapper, it is apparent that the number of highly protected labeling bits also equals $2$.
Moreover, the relationship among AMIs of the four labeling bits can be expressed by $$I(x^{b_1};y)=I(x^{b_2};y)>I(x^{b_4};y)>I(x^{b_3};y).$$
Accordingly, one can utilize a similar method as in {\em Example 1} to protect the VNs at both ends of the entire SC-P codeword.


\begin{figure*}[tp]
\center\vspace{-2mm}
\includegraphics[width=4.5in,height=1.15in]{{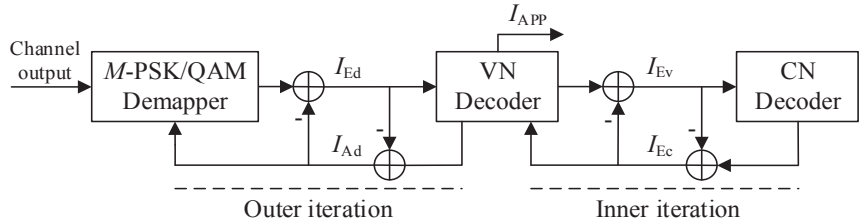}}
\caption{Block diagram of the hierarchical EXIT algorithm for a VNMM-aided SC-P-based BICM-ID system.}
\label{fig:HEXIT}  
\end{figure*}

{\em Remark:}
There exist some other interleaving schemes, e.g., the water-filling interleaving \cite{1605713}, reverse water-filling interleaving \cite{1605713}, and the variable degree matched mapping (VDMM) scheme \cite{5613828,6777400} for LDPC and protograph codes in the existing literature. However, they are tailored for conventional irregular protograph codes, but are not for regular protograph codes. More importantly, these interleaving schemes are not effective for the TB-SC-P codes.

\section{Hierarchical EXIT Algorithm and Performance Analysis for SC-P-based BICM-ID Systems}\label{sect:section-5}

\subsection{Hierarchical EXIT Algorithm}\label{sect:section-C}
In order to accurately evaluate the decoding threshold of SC-P codes in the VNMM-aided BICM-ID systems, a hierarchical EXIT algorithm is proposed in this section.
Specifically, the hierarchical EXIT algorithm can be used to trace the iterative convergence behavior of the MI between the demapper (inner demapper) and the SC-P decoder (outer decoder).
The structure of the hierarchical EXIT algorithm is shown in Fig.~\ref{fig:HEXIT}.
Aiming to facilitate the description of the hierarchical EXIT algorithm process, we define the following concepts.

Let $l \sim {\mathcal N}(\frac{\sigma^2_{{\rm ch}}}{2}, \sigma^2_{{\rm ch}})$ be the LLR of the received signal corresponding to the coded bit over an AWGN channel,
where ${\mathcal N}(\alpha, \beta)$ denotes a Gaussian distribution with mean $\alpha$ and variance $\beta$.
The MI between the coded bit and its corresponding LLR $l$ can be expressed as
\begin{equation}\small
\begin{split}
J(\sigma_{{\rm ch}}) = 1 - \int_{-\infty}^{+\infty} \frac{\rm{exp}\Big({-}\frac{(\mu - \sigma_{{\rm ch}}^2/2)^2}{2\sigma_{{\rm ch}}^2}\Big)}{\sqrt{2 \pi \sigma_{{\rm ch}}^2}}\log_2[1 + \rm{exp}(-\mu)]\,\mathrm{d}\mu.
\end{split}
\end{equation} \normalsize
The derivation of $J(\cdot)$ and its inverse function $J^{-1}(\cdot)$ can be obtained in \cite{1291808}.
The closed-form approximation of $J^{-1} (\cdot)$ is also provided in \cite{1291808}.
Moreover, several different kinds of MIs for the SC-P-based BICM-ID system can be defined as follows.
\begin{itemize}
\item $I_{\rm Ad}$ and $I_{\rm Ed}$ denote the $a$-$priori$ MI and the $extrinsic$ MI of the demapper, respectively.

\item $I_{\rm Av}(i,j)$ denotes the $a$-$priori$ MI transmitted from $c_i$ to $v_j$.

\item $I_{\rm Ac}(i,j)$ denotes the $a$-$priori$ MI transmitted from $v_j$ to $c_i$.

\item $I_{\rm Ev}(i,j)$ denotes the $extrinsic$ MI transmitted from $v_j$ to $c_i$.

\item $I_{\rm Ec}(i,j)$ denotes the $extrinsic$ MI transmitted from $c_i$ to $v_j$.

\item $I_{\rm APP}(j)$ denotes the $a$-$posteriori$ MI of $v_j$.
\end{itemize}


Additionally, we assume that the maximum number of outer iterations between the demapper and the SC-P decoder is set to be $T_{\rm 1}$; while the maximum number of inner iterations for the SC-P decoder in each outer iteration is set to be $T_{\rm 2}$. In particular, we have $I_{\rm Av}(i,j) = I_{\rm Ec}(i,j)$ and $I_{\rm Ac}(i,j) = I_{\rm Ev}(i,j)$ in each inner iteration.
Based on the aforementioned definitions, the hierarchical EXIT algorithm can be described as follows.

\begin{enumerate}[1)\indent]
\item \textbf{Initialization:} At the beginning of the algorithm, there exist an initial $E_b/N_0$ and $m$ initial $a$-$priori$ MIs $I^k_{\rm Ad}$ ($k = 1,2,\ldots,m$), which serve as the input parameters for the demapper.

\item \textbf{Computing the \emph{extrinsic} MI of demapper:} Based on the VNMM-based Monte-Carlo simulation, the $extrinsic$ LLRs corresponding to the coded bits output from the demapper can be obtained. Furthermore, the $extrinsic$ MIs $I^k_{\rm Ed}$ ($k = 1,2,\ldots,m$), which are regarded as the initial decoding information for the VNs in the SC-P decoder, can be computed by
\begin{equation}
\begin{split}
I^k_{\rm Ed}=1-\mathbb{E} \left[ \log_{2} (1+e^{-\widetilde{u}_j l_j})\right],
\end{split}
\end{equation}where $\widetilde{u}_j \in \{+1,-1\}$ denotes the binary-phase shift-keying (BPSK) modulated symbol corresponding to the $j$-th coded bit; $l_j$ denotes the $extrinsic$ LLR of the $j$-th coded bit in the demapper; and $j = \frac{(k-1)n}{m}+1,\frac{(k-1)n}{m}+2,\ldots,\frac{(k-1)n}{m}+\frac{n}{m}$.

\item \textbf{Passing the \emph{extrinsic} MI from demapper to decoder:} Set the initial channel MI of the VN decoder as $I^j_{\rm ch} = I^k_{\rm Ed}$, where $k = 1,2,\ldots,m$ and $j = \frac{(k-1)n_p}{m}+1,\frac{(k-1)n_p}{m}+2,\ldots,\frac{(k-1)n_p}{m}+\frac{n_p}{m}$. As a special case, $I^j_{\rm ch}$ equals zero if the $j$-th VN $v_j$ is punctured in corresponding protograph.

\item \textbf{Updating the \emph{extrinsic} MI from VN to CN:} Exploiting the initial channel MI $I^j_{\rm ch}$ and the $a$-$priori$ MI $I_{\rm Av}(i,j)$ of a VN, one can calculate its corresponding $extrinsic$ MI $I_{\rm Ev}(i,j)$ for $i = 1, 2, \ldots, m_p$ and $j = 1, 2, \ldots, n_p$ (if $b_{i,j} \neq 0$), as
\begin{equation}
\begin{split}
&I_{\rm Ev}(i,j) = J\bigg(\bigg(\sum_{s\neq i}b_{s,j}[J^{-1}(I_{\rm Av}(s,j))]^2\\
&+(b_{i,j}-1)[J^{-1}(I_{\rm Av}(i,j))]^2 + [J^{-1}(I^j_{\rm ch})]^2\bigg)^{1/2}\bigg).
\end{split}
\end{equation} \normalsize

\item \textbf{Updating the \emph{extrinsic} MI from CN to VN:} Similarly, exploiting the $a$-$priori$ MI $I_{\rm Ac}(i,j)$ of a CN, one can also calculate its corresponding $extrinsic$ MI $I_{\rm Ec}(i,j)$, for $i = 1, 2, \cdots, m_p$ and $j = 1, 2, \ldots, n_p$ (if $b_{i,j} \neq 0$), as
\begin{equation}
\begin{split}
I_{\rm Ec}(i,j) =& 1 - J\bigg(\bigg(\sum_{s\neq j}b_{i,s}[J^{-1}(1 - I_{\rm Ac}(i,s))]^2\\
&+(b_{i,j}-1)[J^{-1}(1 - I_{\rm Ac}(i,j))]^2\bigg)^{1/2}\bigg).
\end{split}
\end{equation} \normalsize

\item \textbf{Computing the \emph{a}-\emph{priori} MI of demapper:} Based on the  \emph{a}-\emph{priori} MI $I_{\rm Av}(i,j)$, one can further calculate the average $extrinsic$ MI flowing from the SC-P decoder to the demapper for $k=1,2,\ldots,m$, as
\begin{equation}
\begin{split}
I^k_{\rm Ad} = \frac{1}{(n_p/m)}&\sum\limits_{j=(k-1)n_p/m+1}^{kn_p/m}\\
&J\bigg(\bigg(\sum_{i = 1}^{m_p}b_{i,j}[J^{-1}(I_{\rm Av}(i,j))]^2\bigg)^{1/2}\bigg),
\end{split}
\end{equation} \normalsize which is viewed as updated $a$-$priori$ MIs $I^k_{\rm Ad}$.
\item \textbf{Computing the \emph{a}-\emph{posteriori} MI of VNs:} For $j = 1, 2, \ldots, n_p,$ the $a$-$posteriori$ MI $I_{\rm APP}(j)$ of each VN can be measured exploiting $I_{\rm Av}(i,j)$ and $I^j_{\rm ch}$, as
\begin{equation}
\begin{split}
I_{\rm APP}(j) = J\bigg(\bigg(\sum_{i = 1}^{m_p}b_{i,j}[J^{-1}(I_{\rm Av}&(i,j))]^2\\
&+[J^{-1}(I^j_{\rm ch})]^2\bigg)^{1/2}\bigg).
\end{split}
\end{equation} \normalsize

\item \textbf{Finalization:} The evaluation process is terminated when the $a$-$posteriori$ MI $I_{\rm APP}(j) = 1$ for all $j = 1, 2, \ldots, n_p,$ or when the maximum number of outer iterations is reached. Otherwise, we repeat Steps $2)$-$7)$.
\end{enumerate}

Based on the hierarchical EXIT algorithm, one can get the decoding threshold, i.e., the smallest $E_b/N_0$ that ensures the $a$-$posteriori$ MIs of all VNs in a SC-P code converging to the value of unity.
As a further advancement, Fig.~\ref{fig:new} illustrates the MI update schedule of a protograph-based code in the hierarchical EXIT algorithm.

\emph{Note also that:}
\begin{itemize}
\item The proposed hierarchical EXIT algorithm is a performance-analysis tool for the VNMM-aided SC-P-based BICM-ID systems and the execution process is realized by a serial concatenated scheme, which consists of both Monte-Carlo simulation and information-theoretical derivation. In order to ensure the accuracy of this algorithm, the number of modulated symbols must be sufficiently large.
\item While the proposed hierarchical EXIT algorithm is designed particularly for the VNMM-aided BICM-ID systems, it is also suitable for the BICM systems when the number of outer iteration $T_1$ is set to be $0$.
\item In the standard EXIT algorithm, an infinite-length all-zero codeword can be assumed to be transmitted when calculating the extrinsic MIs due to the characteristics of symmetric channel and asymptotic analysis \cite{7112076,4411526}.
However, the codeword length is finite in practical wireless communication scenarios. In order to obtain a decoding threshold well consistent with the waterfall-region performance
of the simulated BER result, a finite-length binary codeword including both $0$ and $1$ bits is assumed in the proposed hierarchical EXIT algorithm, as in \cite{7080208,1587496,8620336}.

\item In the standard EXIT analysis for the conventional BICM-ID systems, only one $extrinsic$ MI $I_{\rm Ed}$ and one $a$-$priori$ MI $I_{\rm Ad}$ can be obtained in each outer iteration.
Nonetheless, in the proposed hierarchical EXIT analysis, $m$ different $extrinsic$ MIs $I_{\rm Ed}$ and $m$ different $a$-$priori$ MIs $I_{\rm Ad}$ are calculated due to the introduction of the proposed VNMM interleaving scheme. Consequently, the formulas used to calculate the $extrinsic$ MI $I_{\rm Ed}$ and the $a$-$priori$ MI $I_{\rm Ad}$ in the proposed hierarchical EXIT algorithm are different from those in the existing counterparts.

\end{itemize}

\begin{table*}[tp]
\center
\caption{Decoding thresholds of the $(3,6)$ protograph code in BICM-ID systems with different constellation mappers over an AWGN channel. $8$-PSK and $16$-QAM modulations are considered. The maximum numbers of outer iterations $T_{1}$ and inner iterations $T_{2}$ are $8$ and $25$, respectively.}
\begin{tabular}{|c||c|c|c|c|}
\hline
$8$-PSK                     &  SP          & MSEW           &  anti-Gray    & LBPM     \\
\hline
$(E_b/N_0)_{\rm th}$/dB     & $3.049$      & $4.429$        & $4.307$       & $2.480$        \\
\hline\hline
$16$-QAM                    &  SP          & MSEW           &  anti-Gray    & LBPM     \\
\hline
$(E_b/N_0)_{\rm th}$/dB     & $4.724$      & $5.027$        & $4.385$       & $3.626$        \\
\hline
\end{tabular}\label{tab:tab1}
\end{table*}

\begin{table*}[tp]
\center
\caption{Decoding thresholds of the RJA protograph code, $(3,6)$ TB-SC-P code and VNMM-aided $(3,6)$ TB-SC-P code in a BICM-ID system over an AWGN channel. The LBPM-mapped $8$-PSK and $16$-QAM modulations are considered. The maximum numbers of outer iterations $T_{1}$ and inner iterations $T_{2}$ are $8$ and $25$, respectively, and the coupling length for SC-P codes is $12$.}
\begin{tabular}{|c||c|c|c|}
\hline
$8$-PSK                 &  RJA protograph code \cite{6262475}  &  $(3,6)$ TB-SC-P code \cite{7152893}        & VNMM-aided $(3,6)$ TB-SC-P code       \\
\hline
$(E_b/N_0)_{\rm th}$/dB &   $2.363$             &  $2.480$                       & $2.159$                             \\
\hline\hline
$16$-QAM                &  RJA protograph code  &  $(3,6)$ TB-SC-P code          & VNMM-aided $(3,6)$ TB-SC-P code      \\
\hline
$(E_b/N_0)_{\rm th}$/dB &   $3.576$             &  $3.626$                       & $3.410$                             \\
\hline
\end{tabular}\label{tab:tab2}
\end{table*}

\begin{figure}[tpb]
\centering
\subfigure[\hspace{-0.26cm}]{ 
\includegraphics[width=0.19in,height=2.0in]{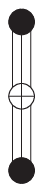}}~~~~~~
\subfigure[\hspace{-0.3cm}]{ 
\includegraphics[width=2.8in,height=2in]{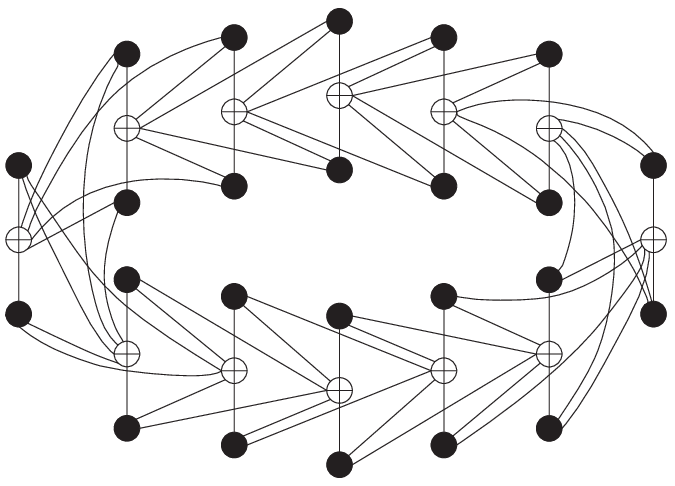}}
\caption{Structures of (a) $(3,6)$ protograph code and (b) $(3,6)$ TB-SC-P code with coupling length $L=12$.}\vspace{-2mm}
\label{fig:Protograph}  
\end{figure}

\subsection{Asymptotic-Performance Analysis}
Considering a specific example, Fig.~\ref{fig:Protograph} shows the protographs of a $(3,6)$ protograph code and its corresponding TB-SC-P code with coupling length $L=12$, where the VNs and CNs are represented by black circles and circles with a plus sign, respectively.
Referring to Fig.~\ref{fig:Protograph}(b), the number of VNs in the protograph of $(3,6)$ TB-SC-P code is $24$.
Through a ``copy-and-permute" operation with $n/24$ copies of this protograph, one can obtain a $(3,6)$ TB-SC-P code with a codeword length of $n$.

Based on the $8$-PSK and $16$-QAM modulations,
the decoding-wave phenomenon and decoding threshold of the $(3,6)$ TB-SC-P code are analyzed by exploiting the hierarchical EXIT algorithm in Sect.~\ref{sect:section-4}.
In order to validate the superiority of the proposed LBPM constellation mappers and VNMM scheme in the BICM-ID condition,
the SP, MSEW, and anti-Gray constellation mappers are used as benchmarks. As a further advance, we also compare the decoding thresholds of the
proposed VNMM-aided $(3,6)$ TB-SC-P code and the well-performing protograph code (i.e.,  repeat-jagged-accumulate (RJA) protograph code) \cite{6262475}.

\subsubsection{Decoding Wave}
\begin{figure}[tbp]
\centering\vspace{-1mm}
\subfigure[\hspace{-0.5cm}]{ 
\includegraphics[width=3.0in,height=2.3in]{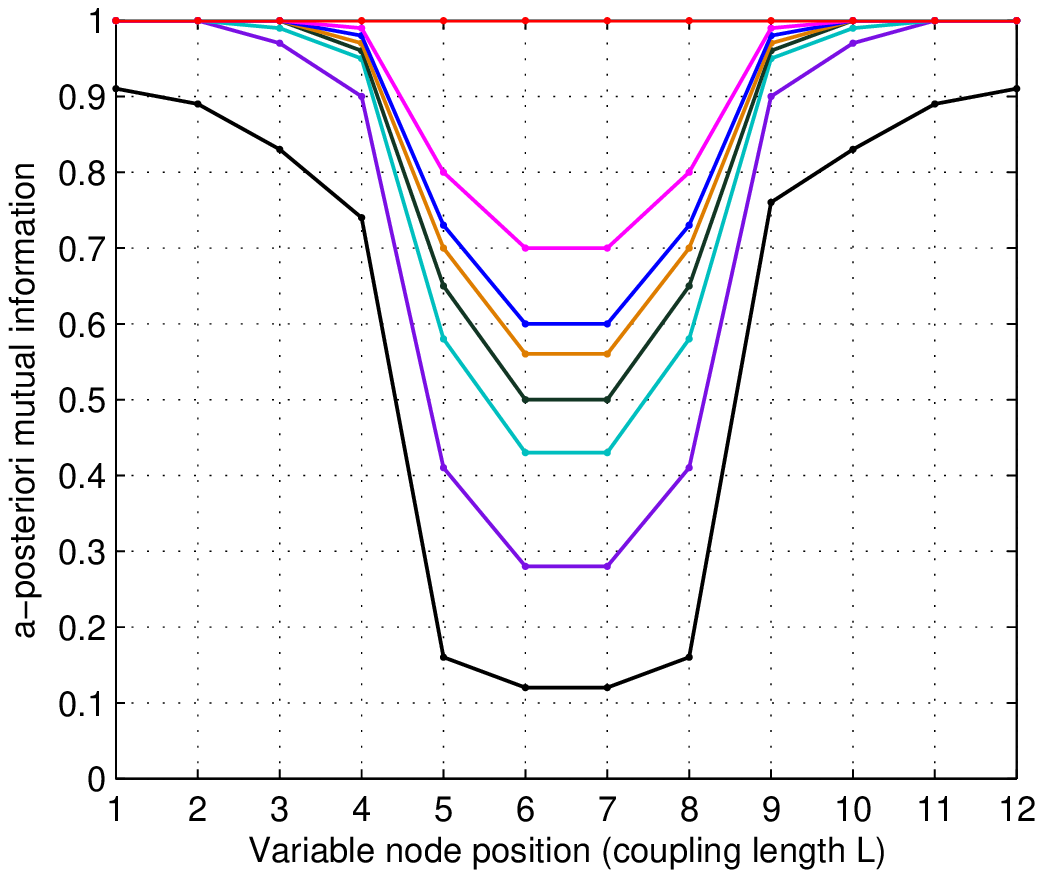}}
\subfigure[\hspace{-0.5cm}]{ 
\includegraphics[width=3.0in,height=2.3in]{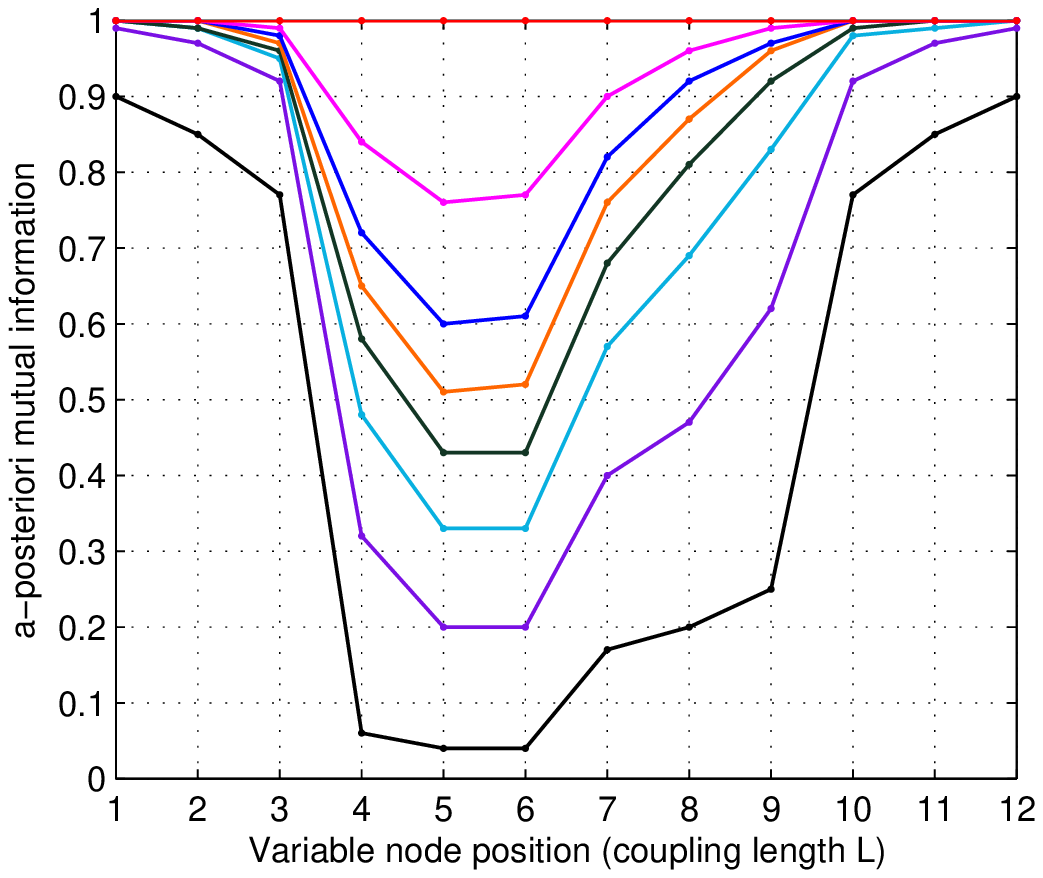}}
\caption{Decoding waves of the $(3,6)$ TB-SC-P code with the proposed LBPM constellation mappers and VNMM scheme in a BICM-ID system over an AWGN channel. The coupling length is $12$ and the outer iterations are $T_{1}=1,2,\ldots,8$ (from bottom to top): (a) $8$-PSK and (b) $16$-QAM modulations.}
\label{fig:side:wave}  
\end{figure}

Fig.~\ref{fig:side:wave} depicts the decoding-wave phenomenon of the $(3,6)$ TB-SC-P code with coupling length $L=12$ based on the proposed LBPM-mapped $8$-PSK and $16$-QAM modulations and the VNMM in a BICM-ID system.
In this figure, we plot the evolution of the a-posteriori MIs corresponding to the VNs in coupling position $(L=1, 2, \ldots, 12)$
as the number of outer iterations $T_1$ ($T_1=1,2,\ldots,8$) increases.
As seen, the a-posteriori MIs for the VNs at both ends of the protograph can quickly converge to the value of one with an increase of the number of outer iterations.
Moreover, a wave-like convergence from both ends to the middle area is yielded as the a-posteriori MIs continuously increases.
From the viewpoint of iterative decoding, the VNs at both ends of the protograph can pass more reliable messages to their neighboring VNs.

\begin{figure*}[tbp]
\centering\vspace{-3mm}
\subfigure[\hspace{-0.5cm}]{ 
\includegraphics[width=3.0in,height=2.3in]{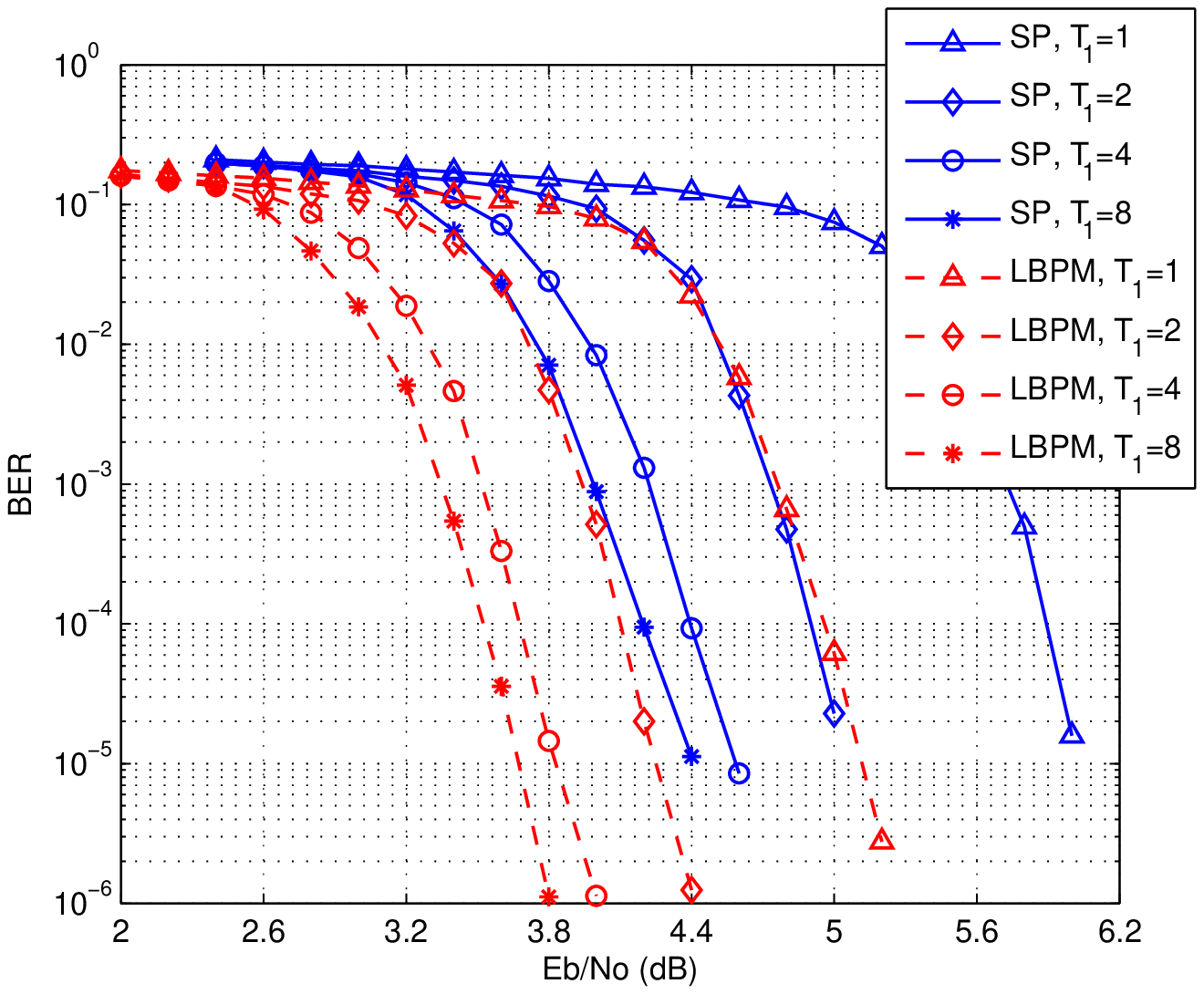}}
\subfigure[\hspace{-0.5cm}]{ 
\includegraphics[width=3.0in,height=2.3in]{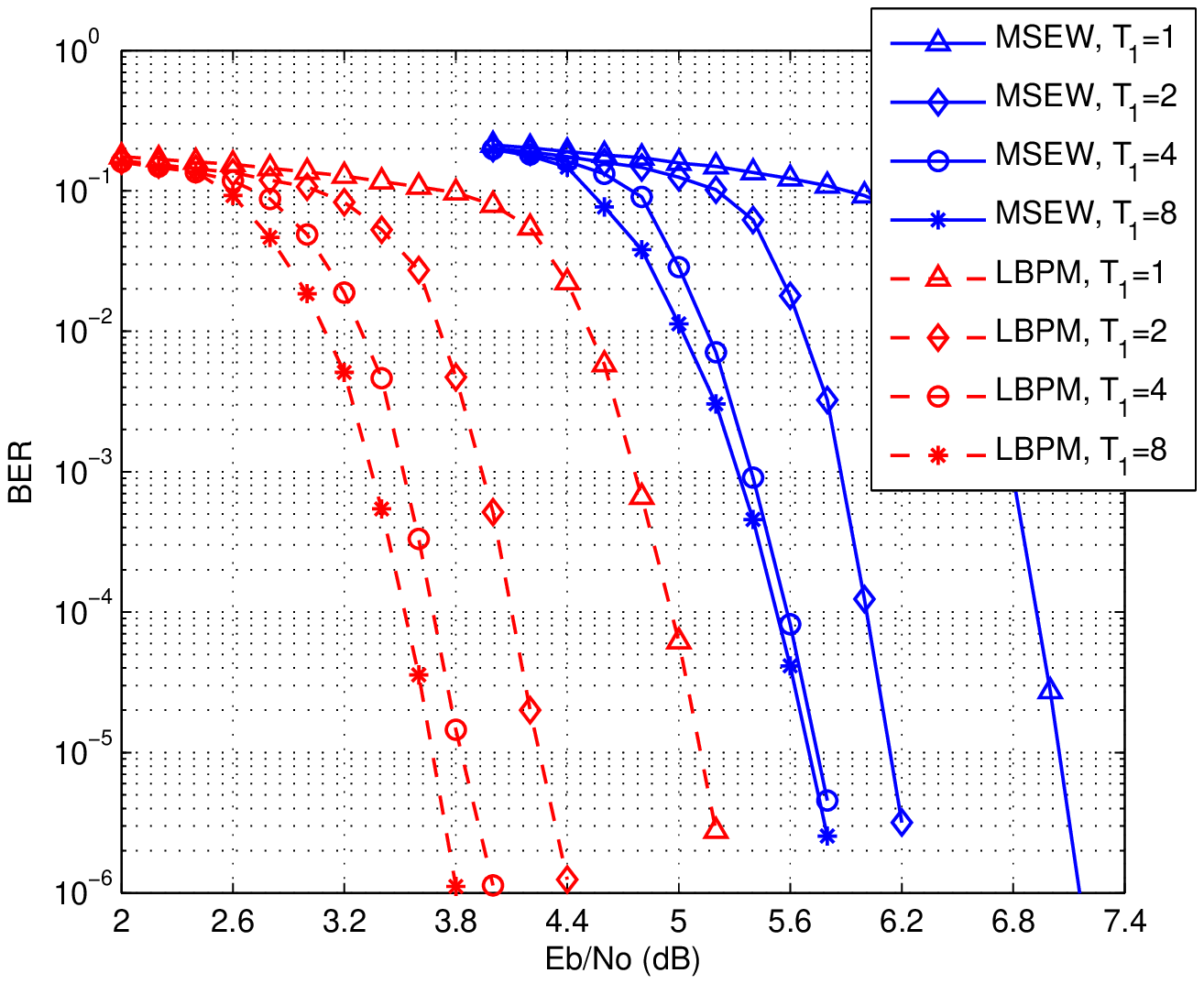}}
\subfigure[\hspace{-0.5cm}]{ 
\includegraphics[width=3.0in,height=2.3in]{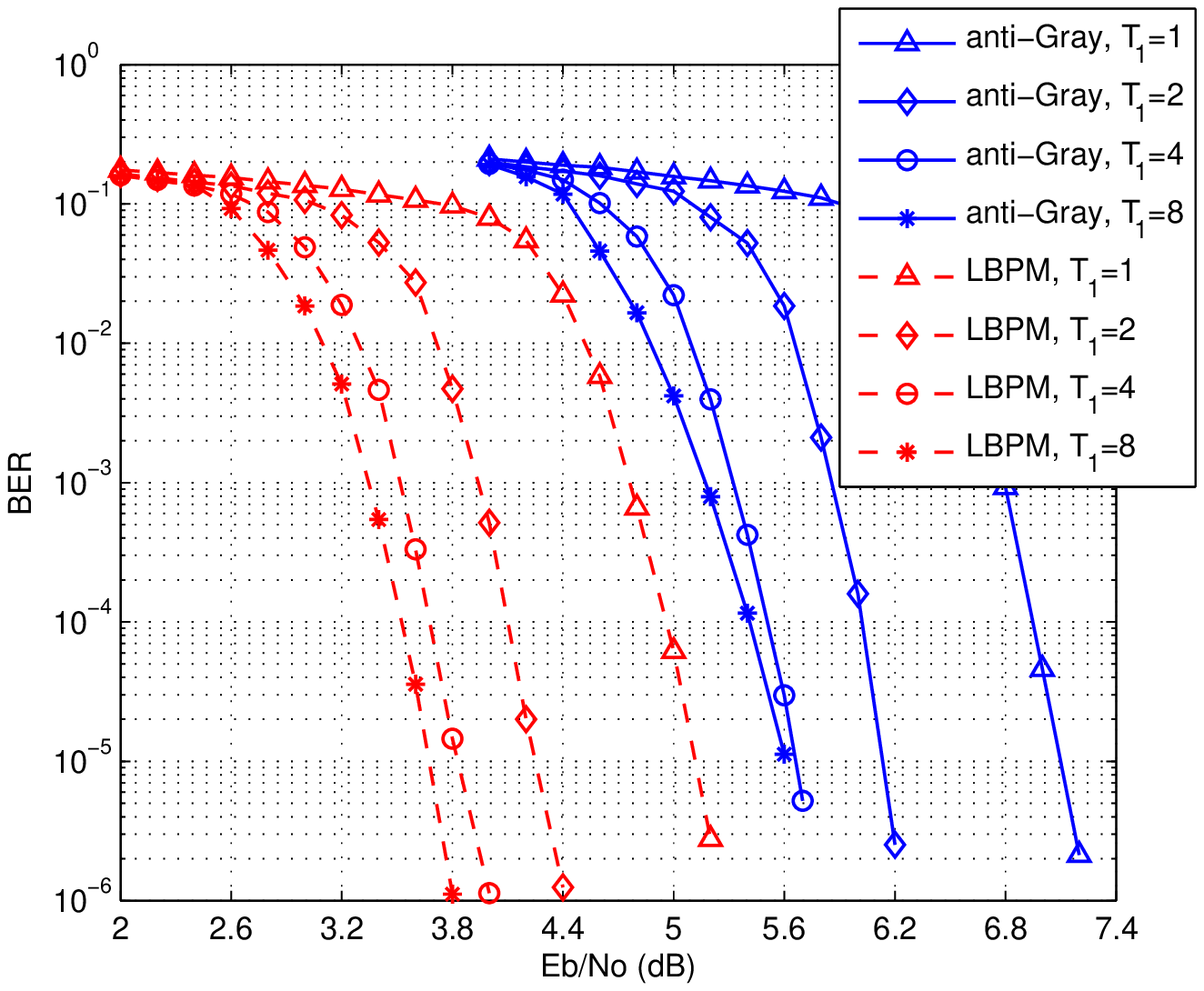}}
\subfigure[\hspace{-0.5cm}]{ 
\includegraphics[width=3.0in,height=2.3in]{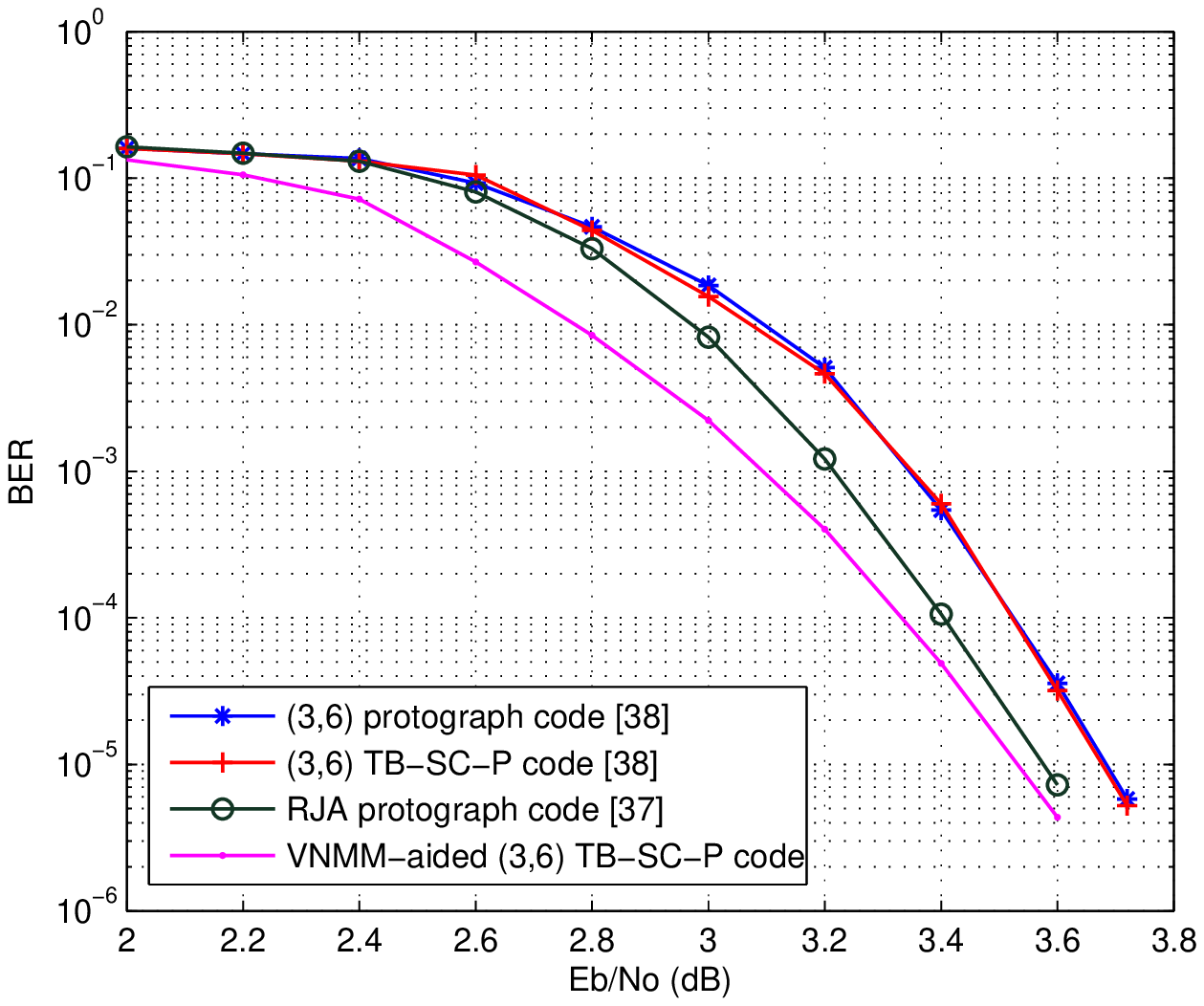}}
\caption{BER curves of the $(3,6)$ protograph code in the BICM-ID systems with different constellation mappers: (a) SP and LBPM constellation mappers, (b) MSEW and LBPM constellation mappers, (c) anti-Gray and LBPM constellation mappers; and (d) BER curves of the $(3,6)$ protograph code, $(3,6)$ TB-SC-P code, RJA protograph code, and VNMM-aided $(3,6)$ TB-SC-P code in a BICM-ID system with LBPM constellation mapper, where the maximum numbers of outer iterations $T_1$ and inner iterations $T_2$ are $8$ and $25$, respectively, and the coupling length for the TB-SC-P codes is $12$. An $8$-PSK modulation is considered.}\vspace{-2mm}
\label{fig:side:8PSK-anti-Gray-maximizing}  
\end{figure*}

\begin{figure*}[tpb]
\centering\vspace{-3mm}
\subfigure[\hspace{-0.5cm}]{ 
\includegraphics[width=3.0in,height=2.3in]{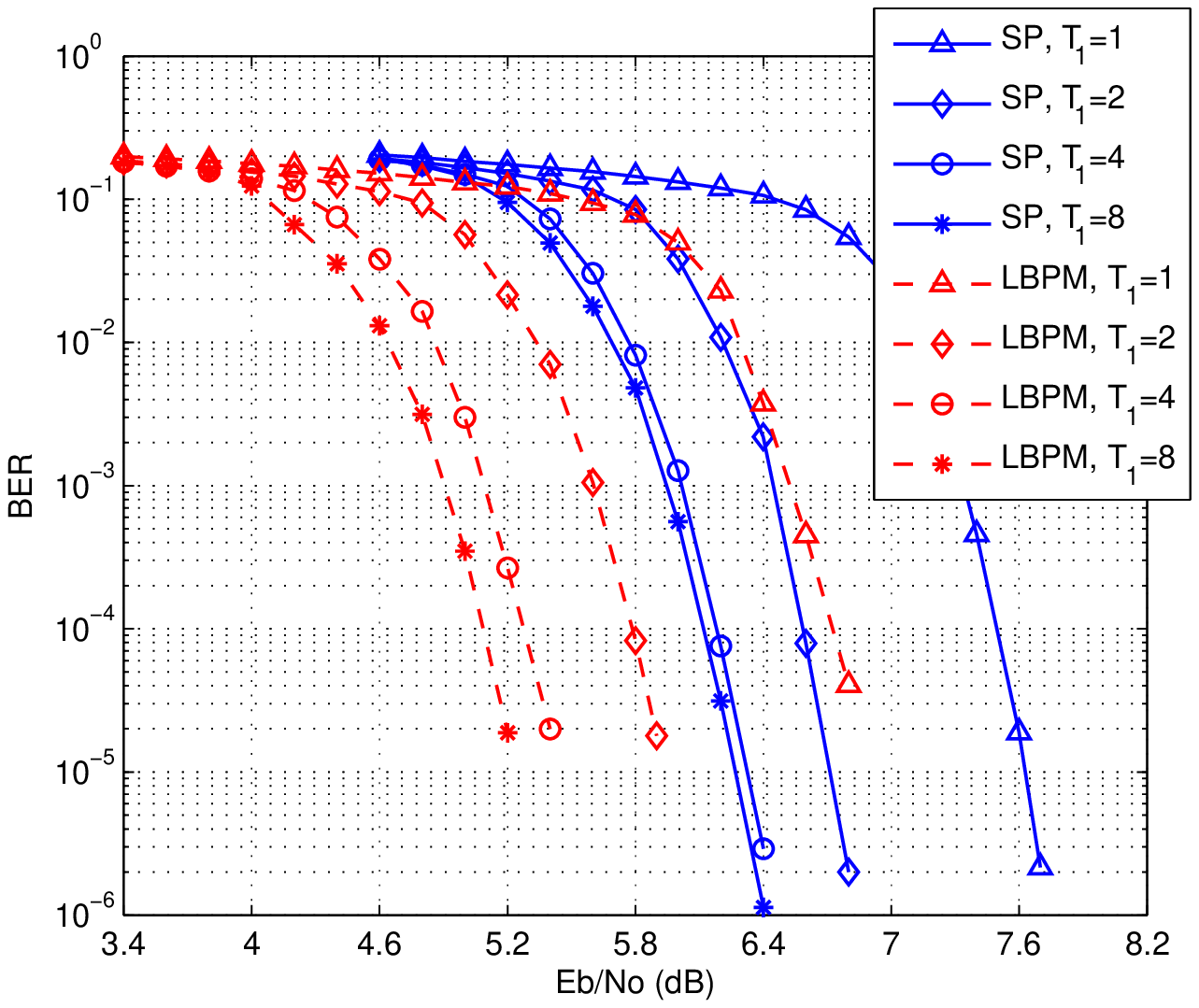}}
\subfigure[\hspace{-0.5cm}]{ 
\includegraphics[width=3.0in,height=2.3in]{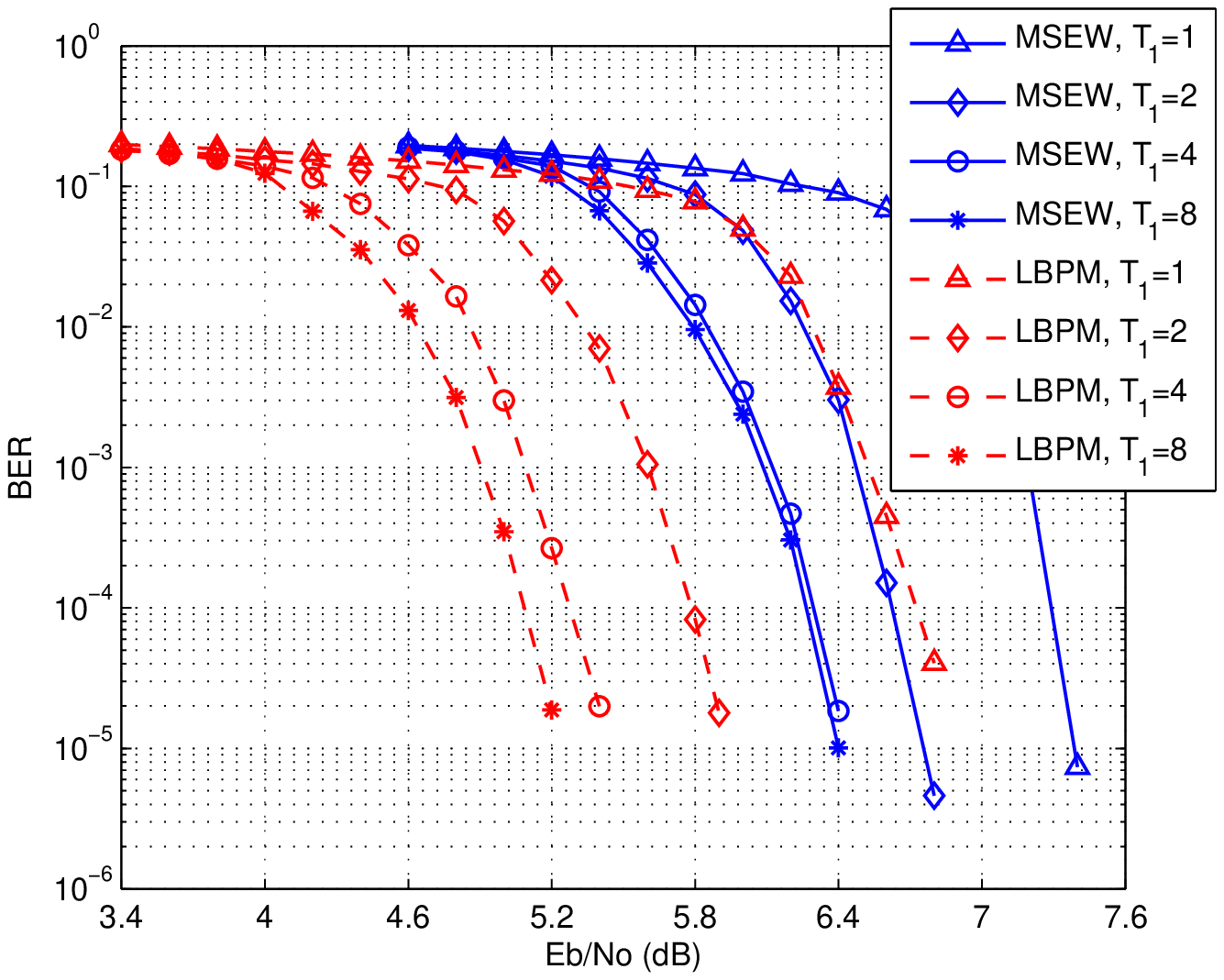}}
\subfigure[\hspace{-0.5cm}]{ 
\includegraphics[width=3.0in,height=2.3in]{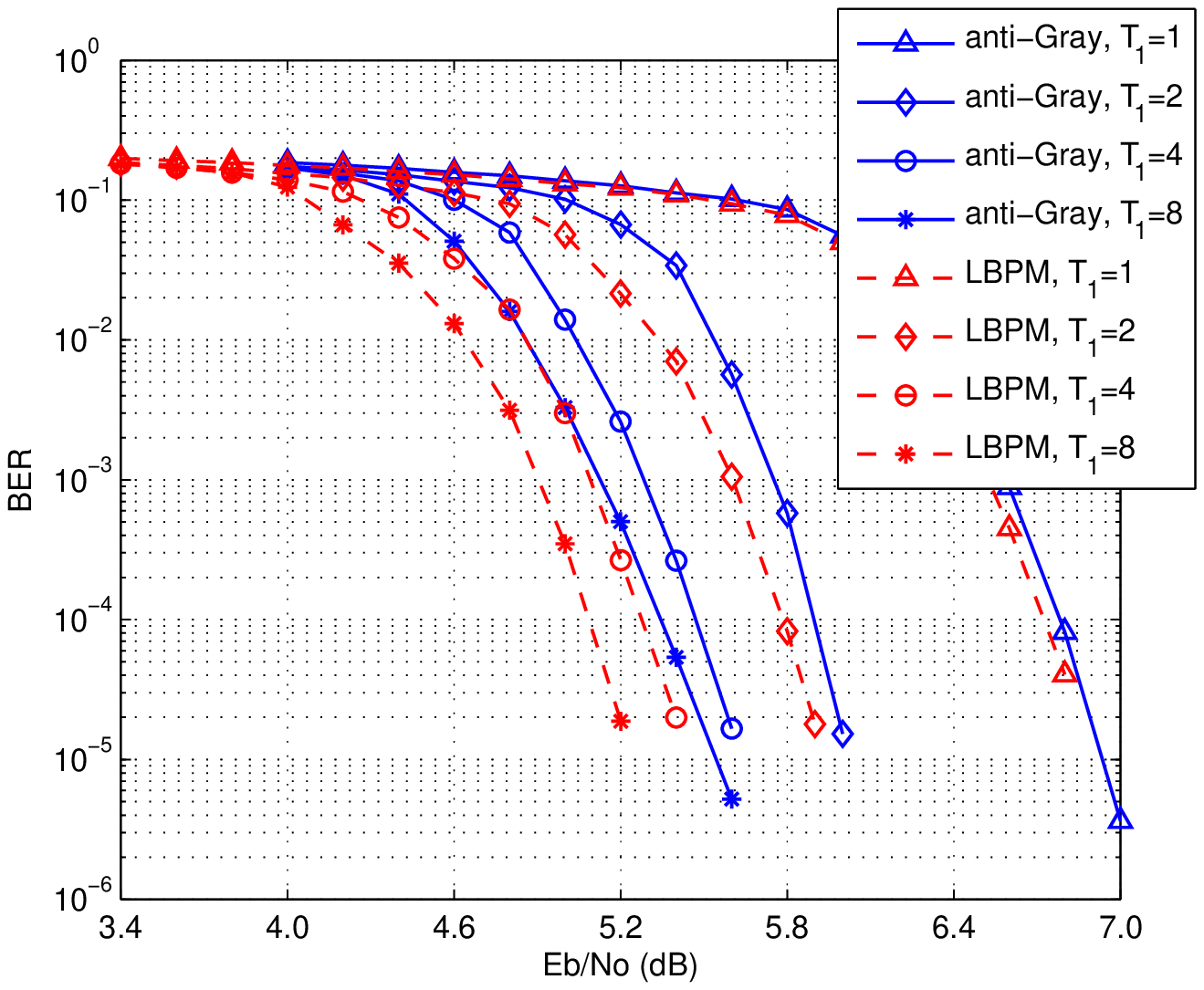}}
\subfigure[\hspace{-0.5cm}]{ 
\includegraphics[width=3.0in,height=2.3in]{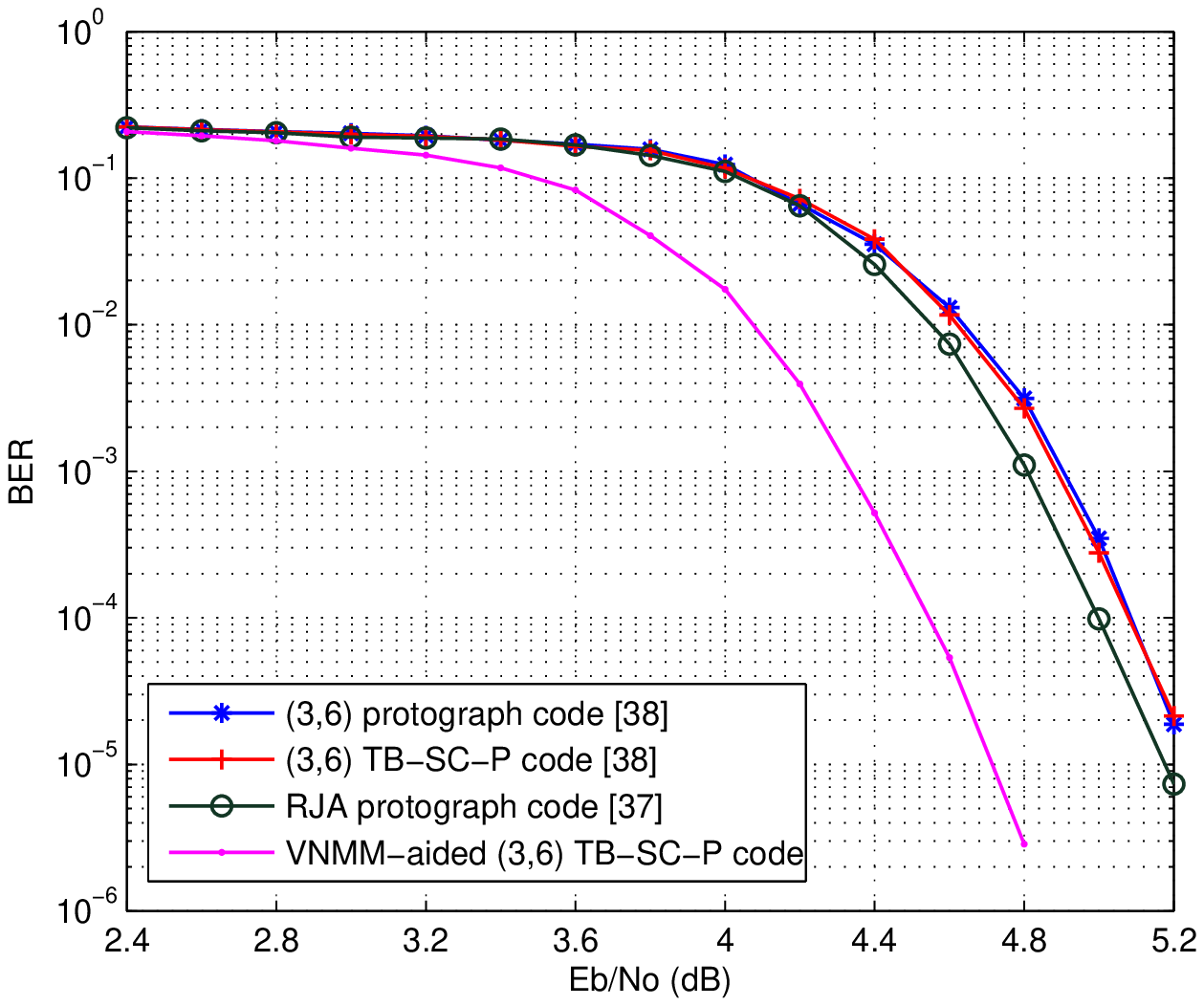}}
\caption{BER curves of the $(3,6)$ protograph code in the BICM-ID systems with different constellation mappers: (a) SP and LBPM constellation mappers, (b) MSEW and LBPM constellation mappers, (c) anti-Gray and LBPM constellation mappers; and (d) BER curves of the $(3,6)$ protograph code, $(3,6)$ TB-SC-P code, RJA protograph code, and VNMM-aided $(3,6)$ TB-SC-P code in a BICM-ID system with LBPM constellation mapper, where the maximum numbers of outer iterations $T_1$ and inner iterations $T_2$ are $8$ and $25$, respectively, and the coupling length for the TB-SC-P codes is $12$. A $16$-QAM modulation is considered.}\vspace{-2mm}
\label{fig:side:16QAM-anti-Gray-maximizing}  
\end{figure*}

\subsubsection{Decoding Threshold}

Based on different constellation mappers, we estimate the decoding thresholds of
the original $(3,6)$ protograph code in BICM-ID systems by exploiting the hierarchical EXIT algorithm, and present the results in Table~\ref{tab:tab1}, where $8$-PSK and $16$-QAM modulations are considered.
As observed, the $(3,6)$ protograph code with the proposed LBPM constellation mappers can exhibit the smallest decoding thresholds compared with the other three existing constellation mappers.
Considering the proposed LBPM constellation mappers,
the MIs of $(3,6)$ protograph code can converge to the value of unity at $E_b/N_0 =2.480$~{dB} and $E_b/N_0 =3.626$~{dB} for $8$-PSK and $16$-QAM modulations, respectively, while the $(3,6)$ protograph code with other constellation mappers require much larger signal-to-noise ratios (SNRs).
This implies that the $(3,6)$ protograph code can achieve the best error performance when the proposed LBPM constellation mappers are applied in the BICM-ID systems.

In order to get further insight, we compare in Table~\ref{tab:tab2} the decoding thresholds of
(i) the RJA protograph code \cite{6262475},
(ii) the $(3,6)$ TB-SC-P code \cite{7152893}, and (iii) the VNMM-aided $(3,6)$ TB-SC-P code in BICM-ID systems with the proposed LBPM constellation mappers. Note that the $(3,6)$ TB-SC-P code
 in Table~\ref{tab:tab2} has the same decoding thresholds (i.e., $2.480$~dB and $3.626$~dB) as the $(3,6)$ protograph code in Table~\ref{tab:tab1}.
More importantly, the decoding threshold of the $(3,6)$ TB-SC-P code can be further improved by utilizing the VNMM scheme.
Moreover, the VNMM-aided $(3,6)$ TB-SC-P code can achieve an additional threshold gain with respect to the RJA protograph code, which outperforms the $(3,6)$ TB-SC-P code and $(3,6)$ protograph code.
Hence, the SC-P-based BICM-ID system with the proposed LBPM constellation mappers and VNMM scheme is able to achieve excellent error performances.

{\em Remark:} To illustrate the potential
benefits of the LBPM constellation mapper and VNMM scheme in a simple and clear way, we mainly focus on the
regular $(3, 6)$ SC-P code and PSK/QAM in this paper. However, the proposed design and analysis methodologies are applicable to other complex-valued modulations and other SC-P codes.
%
%

\section{Simulation Results}
Based on the BICM-ID systems, we present various simulation results of the $(3,6)$ protograph code, RJA protograph code,
and $(3,6)$ TB-SC-P code over AWGN channels so as to demonstrate the merit of the proposed LBPM constellation mappers and VNMM scheme.
In particular, the code rate and codeword length are assumed to be $1/2$ and $4800$.
Furthermore, the number of outer iterations between the demapper and the protograph/SC-P decoder is set to be $T_{1}=1,2,4,8$,
while the maximum number of inner iterations is set to be $T_{2}=25$.


\subsection{BER Performance for 8-PSK Modulated BICM-ID Systems}


For the $8$-PSK modulation, the bit-error-rate (BER) curves of the $(3,6)$ protograph code in the BICM-ID systems with different constellation mappers are depicted in Fig.~\ref{fig:side:8PSK-anti-Gray-maximizing}.
As observed from Fig.~\ref{fig:side:8PSK-anti-Gray-maximizing}(a)-(c),
the $(3,6)$ protograph code with the proposed LBPM constellation mapper can exhibit the best error performance
compared with the conventional constellation mappers (i.e., the SP, MSEW, and anti-Gray constellation mappers)
 in the BICM-ID systems.
For the case of $T_1=8$,
the $(3,6)$ protograph code with the proposed LBPM constellation mapper only requires an SNR of about $3.7~{\rm dB}$ to accomplish a BER of $10^{-5}$,
while the codes with SP, anti-Gray, and MSEW constellation mappers require SNRs of about $4.4~{\rm dB}$, $5.6~{\rm dB}$, $5.7~{\rm dB}$, respectively.
In other words, the proposed LBPM constellation mapper possesses remarkable gains of about $0.7~{\rm dB}$, $1.9~{\rm dB}$ and $2.0~{\rm dB}$ over the SP, anti-Gray, and MSEW constellation mappers, respectively, at a BER of $10^{-5}$.
The above simulated results are reasonably consistent with our theoretical analyses in Sect.~V-B.
In addition, as can be seen, the proposed LBPM constellation mapper can attain a gain of about $0.2~{\rm dB}$
when the number of outer iterations increases from $4$ to $8$.
In this sense, an additional gain can be expected for the proposed LBPM constellation mapper when the number of outer iterations becomes even larger.


\begin{figure}[tbp]
\centering\vspace{-1mm}
\subfigure[\hspace{-0.5cm}]{ 
\includegraphics[width=3.0in,height=2.3in]{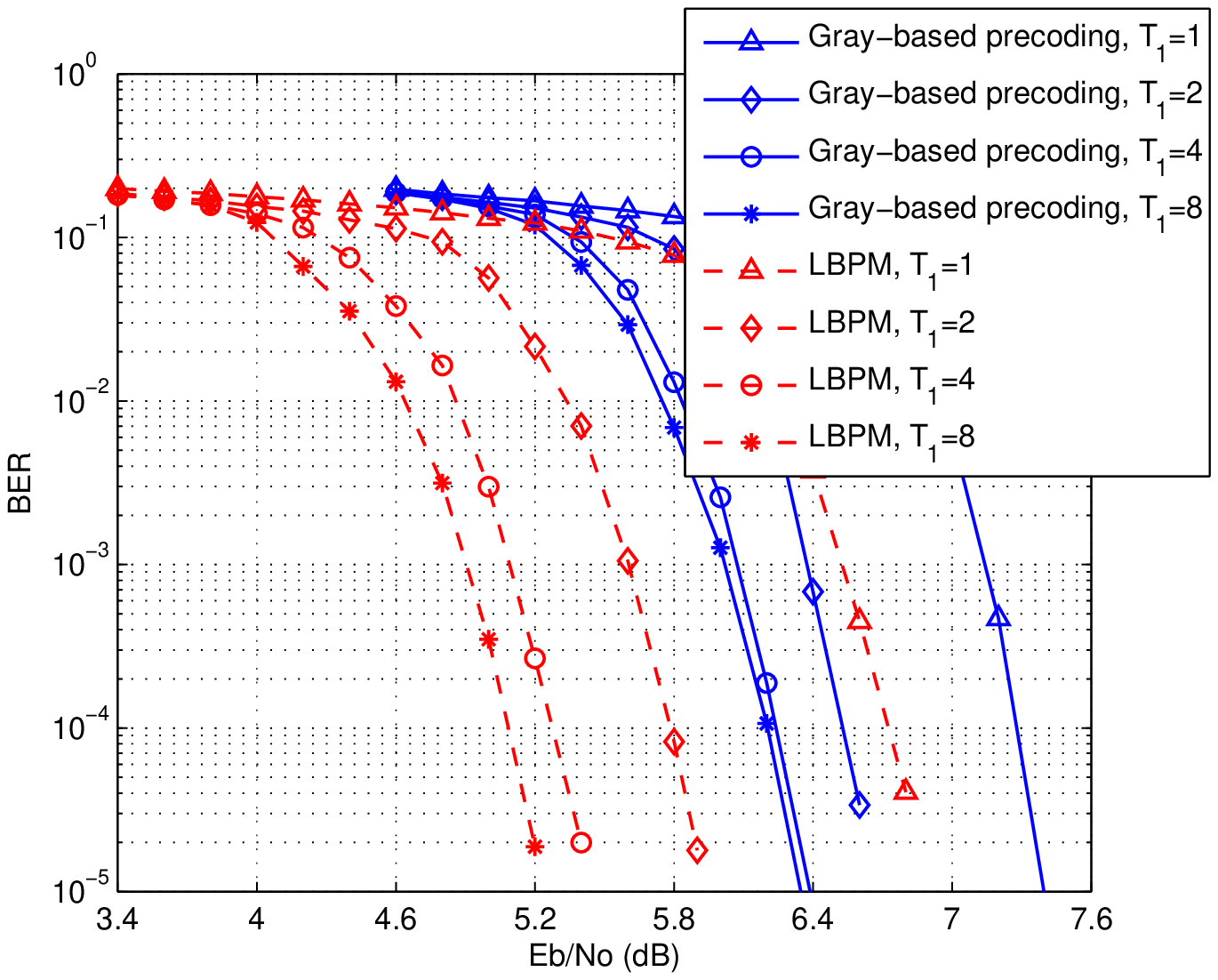}}
\subfigure[\hspace{-0.5cm}]{ 
\includegraphics[width=3.0in,height=2.3in]{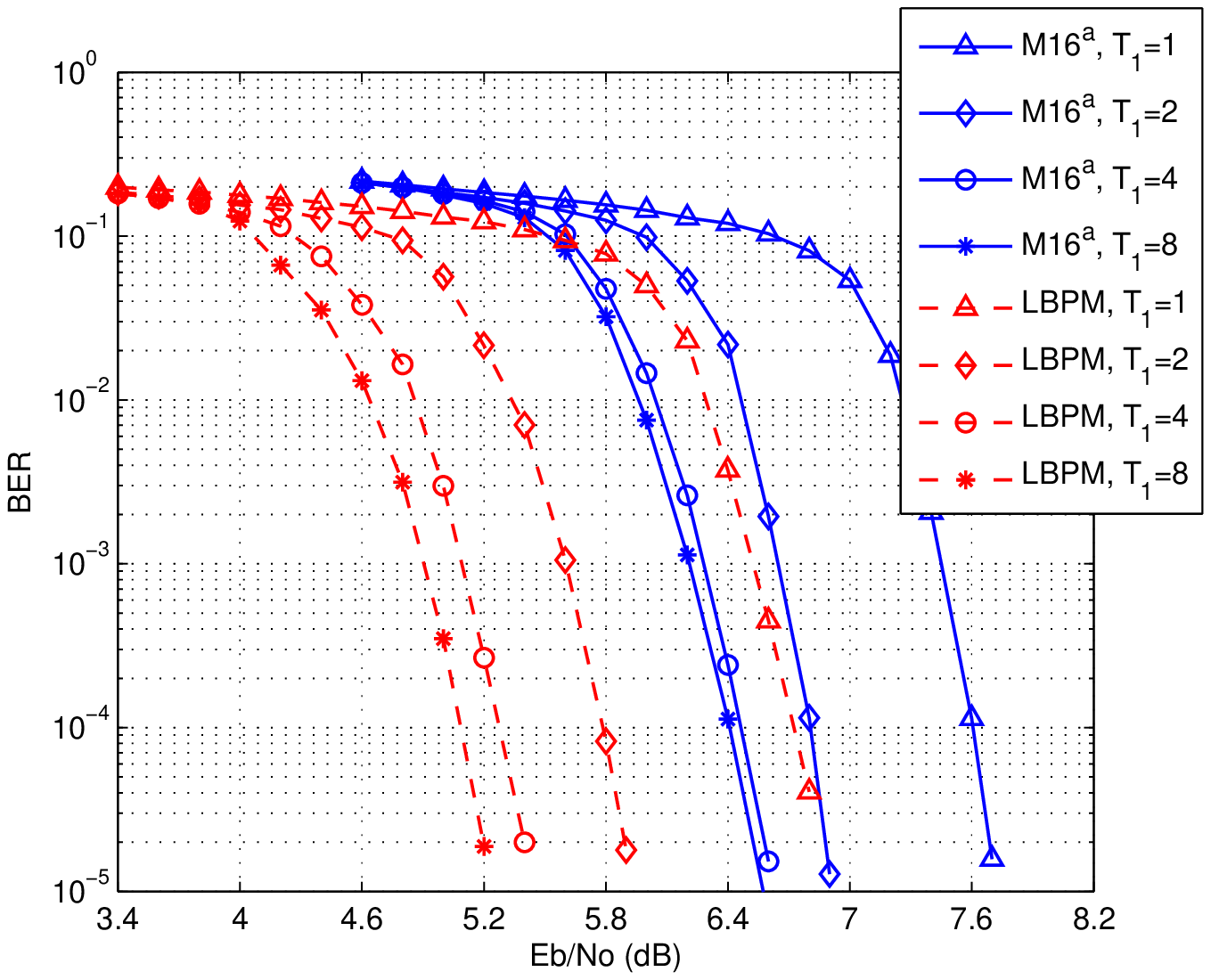}}
\caption{BER curves of the $(3,6)$ protograph code in the BICM-ID systems with different constellation mappers: (a) Gray-based precoding \cite{6823707} and LBPM constellation mappers, (b) $M16^a$ \cite{1254064} and LBPM constellation mappers. A $16$-QAM modulation is considered.}
\label{fig:compare}
\end{figure}

In order to validate the superiority of the VNMM scheme in SC-P-based BICM-ID systems, Fig.~\ref{fig:side:8PSK-anti-Gray-maximizing}(d) presents the BER curves of the $(3,6)$ protograph code \cite{7152893}, $(3,6)$ TB-SC-P code \cite{7152893}, RJA protograph code \cite{6262475}, and the VNMM-aided (3,6) TB-SC-P code in a BICM-ID system with the proposed LBPM-mapped $8$-PSK modulation.
As can be observed, the error performance of the VNMM-aided $(3,6)$ TB-SC-P code is much better than its
corresponding $(3,6)$ TB-SC-P code, which has the same performance as the originial $(3,6)$ protograph code.
Specifically, the VNMM-aided $(3,6)$ TB-SC-P code can achieve a gain of more than $0.2$ dB over the $(3,6)$ TB-SC-P code in the whole BER range under study, which demonstrates the merit of our interleaver design.
Furthermore, the VNMM-aided $(3,6)$ TB-SC-P code outperforms the well-performing RJA protograph code, especially in the low SNR region. The performance gap between these two types of codes becomes smaller in the high SNR region due to the fact that the unequal
protection-degree property cannot be substantially exploited when the modulation order is too small ($M \le 8$).
In fact, the performance benefit of the proposed VNMM scheme will be more noticeable for higher-order modulations (see Fig.~\ref{fig:side:16QAM-anti-Gray-maximizing}(d)).


\subsection{BER Performance for 16-QAM Modulated BICM-ID Systems}
Similarly, Fig.~\ref{fig:side:16QAM-anti-Gray-maximizing}(a)-(c) show the BER curves of the $(3,6)$ protograph code in the BICM-ID systems with different constellation mappers, when a $16$-QAM modulation is considered.
As can be seen, the proposed LBPM constellation mapper allows the $(3,6)$ protograph code to exhibit the best performance compared with the other three conventional constellation mappers. Generally speaking, the relative performance among the proposed LBPM, SP, MSEW, and anti-Gray constellation mappers in the case of $16$-QAM modulation is identical to that in the case of $8$-PSK modulation. This phenomenon verifies the effectiveness of the LBPM constellation mapper in the scenarios with different modulation orders.


Based on the LBPM-mapped $16$-QAM modulation, we also compare the BER performance of four different protograph codes in the BICM-ID systems and present the results in Fig.~\ref{fig:side:16QAM-anti-Gray-maximizing}(d).
Different from the results in the $8$-PSK modulation, the VNMM-aided $(3,6)$ TB-SC-P code achieves a significant gain of about $0.4~{\rm dB}$ over the RJA protograph code at a BER of $10^{-5}$, which further achieves a gain of about $0.1~{\rm dB}$ over the $(3,6)$ TB-SC-P code and the original $(3,6)$ protograph code.
According to the above observations, the proposed interleaving scheme not only significantly improves the SC-P-based BICM-ID systems, but also makes the SC-P codes a competitive alternative to other ECCs in high-rate and high-reliability wireless communication applications.

To further verify the superiority of the proposed LBPM constellation mappers, another two types of state-of-the-art counterparts (i.e., referred to as {\em Gray-based precoding constellation mapper} \cite{6823707} and {\em $M16^a$ constellation mapper} \cite{1254064}) are used as benchmarks for performance comparison. As observed from Fig.~\ref{fig:compare}(a), the $(3,6)$ protograph code with the proposed LBPM constellation mapper can exhibit better error performance than that with the Gray-based precoding constellation mapper. For the case of $T_1 = 8$, the $(3,6)$ protograph code with the proposed LBPM constellation mapper only requires
an SNR of about $5.2$ dB to accomplish a BER of $2 \times 10^{-5}$, while the code with the Gray-based precoding constellation mapper requires an SNR of about $6.3$ dB to do so.
Thus, the proposed LBPM constellation mapper possesses a gain of about $1.1$ dB over the Gray-based precoding constellation mapper at a BER of $2 \times 10^{-5}$. Similar conclusions can be drawn from
Fig.~\ref{fig:compare}(b) on the $M16^a$ constellation mapper.

\begin{figure}[tbp]\vspace{-2mm}
\center
\includegraphics[width=3.0in,height=2.3in]{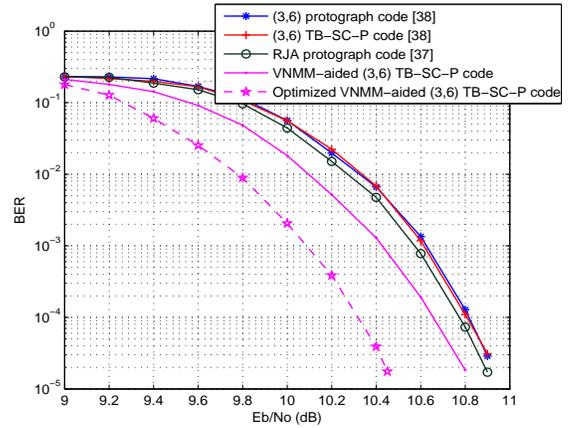}
\vspace{-3mm}
\caption{BER curves of the $(3,6)$ protograph code, $(3,6)$ TB-SC-P code, RJA protograph code, VNMM-aided $(3,6)$ TB-SC-P code, and optimized VNMM-aided $(3,6)$ TB-SC-P code in a BICM-ID system. The LBPM-mapped $64$-QAM modulation is considered. The maximum numbers of outer iterations $T_1$ and inner iterations $T_2$ are $8$ and $25$, respectively, and the coupling length for the SC-P codes is $12$.}
\label{fig:Fig.64QAM}
\end{figure}

\subsection{BER Performance for 64-QAM Modulated BICM-ID Systems}
As a final example, we further perform simulations on the original $(3,6)$ protograph code, $(3,6)$ TB-SC-P code, RJA protograph code, VNMM-aided $(3,6)$ TB-SC-P code, and optimized VNMM-aided $(3,6)$ TB-SC-P code in a BICM-ID system with the LBPM-mapped $64$-QAM modulation in Fig.~\ref{fig:Fig.64QAM} to demonstrate the feasibility of the VNMM scheme in higher-order-modulation scenarios.\footnote{The principle of optimized VNMM scheme will be explained at the end of this paragraph.}
It should be noted that the LBPM constellation mapper for the $64$-QAM modulation used here can be easily constructed by exploiting the two-step design method mentioned in Sect.~\ref{sect:section-3}. As expected, our proposed VNMM-aided $(3,6)$ TB-SC-P code can preserve its advantage and is still superior to the RJA protograph code, $(3,6)$ TB-SC-P code, and original $(3,6)$ protograph code in terms of BER performance.
Nonetheless, the performance gain from the VNMM scheme under the LBPM-mapped $64$-QAM modulation is smaller than that under the LBPM-mapped $16$-QAM modulation.
Specifically, at a BER of $2 \times 10^{-5}$, the VNMM-aided $(3,6)$ TB-SC-P code achieves a gain of about $0.4~{\rm dB}$ over the RJA protograph code in the case of $16$-QAM modulation, while the former only achieves a gain of about $0.1$~dB over the latter in $64$-QAM modulation.
This phenomenon is due to the fact that the number of highly protected labeling bit positions is set as $2$ in the proposed VNMM scheme.
Therefore, although the decoding-wave phenomenon can be trigged in $64$-QAM modulation, the number of high-priority blocks is only $50\%$ of that of the general-priority blocks (i.e., the numbers of high-priority blocks and general-priority blocks equal $2$ and $4$, respectively).\footnote{The block whose coded bits are assigned to the highly protected labeling bit positions is defined as a high-priority block, while the block whose coded bits are assigned to the generally protected labeling bit positions is defined as a general-priority block.} Thus, the coded bits in the middle area of the codeword in this case cannot converge as fast as those in the case of $16$-QAM modulation.

To address this problem, the proposed VNMM scheme can be optimized according to the value of modulation order $M$ (especially for $M \ge 64$) as follows. The number of protection degrees can be set as $\lceil m/2 \rceil$ rather than $2$, where $\lceil r \rceil$ denotes the ceiling $r$. In other words, $m$ labeling bit positions can be divided into $\lceil m/2 \rceil$ different priorities. Subsequently, the $m$ blocks of a given codeword from the both ends to the middle area are assigned to the highest-priority, the second highest-priority, \ldots, the lowest-priority labeling bit positions, respectively.
In order to verify the effectiveness of the optimized VNMM scheme, we apply such an optimized scheme to the LBPM-mapped $64$-QAM modulation and include the corresponding BER curve in Fig.~\ref{fig:Fig.64QAM}.
As can be seen, the optimized VNMM-aided $(3,6)$ TB-SC-P code achieves a significant gain of about $0.4$~dB over the (original) VNMM-aided $(3,6)$ TB-SC-P code at a BER of $2\times10^{-5}$.


\section{Conclusions}
In this paper, we have investigated the design and performance of spatially-coupled protograph (SC-P)-based BICM-ID systems, in which $M$-ary~$(M \ge8)$ PSK and QAM modulations are adopted. In particular, we put forward a two-step design approach to construct a type of LBPM constellation mappers for the SC-P-based BICM-ID systems. The proposed LBPM constellation mappers not only can allow the BICM systems to exhibit desirable capacities, but also can ensure excellent iterative performance of the BICM-ID systems.
Furthermore, we have conceived a novel interleaving scheme, called {\em VNMM scheme}, tailored for use in SC-P-based BICM-ID systems to trigger the decoding-wave phenomenon in the turbo-like decoding process, which can therefore accelerate the convergence of a-posteriori MI corresponding to the TB-SC-P codes.
As a further advancement, we have conceived a hierarchical EXIT algorithm to estimate the asymptotic decoding thresholds of TB-SC-P codes in the VNMM-aided BICM-ID systems.
Both hierarchical EXIT analyses and simulations have illustrated that the proposed LBPM-VNMM-aided SC-P-based BICM-ID systems can achieve very desirable error performance and can significantly outperform the conventional BICM-ID systems. Thanks to the performance superiority, the proposed BICM-ID systems appear to be a good transmission technique for future wireless communication applications.


\end{document}